\documentclass{article}
\usepackage{amssymb}
\usepackage{amsmath}
\usepackage{longtable}
\usepackage{booktabs}
\usepackage[ruled,linesnumbered]{algorithm2e}
\usepackage{multirow,bigstrut,bigdelim}
\usepackage{authblk}

\newtheorem{Thm}{Theorem}[section]

\newtheorem{Def}[Thm]{Definition}
\newtheorem{Pro}[Thm]{Proposition}
\newtheorem{La}[Thm]{Lemma}
\newtheorem{Rem}[Thm]{Remark}

\newtheorem{Ex}[Thm]{Example}

\newenvironment{Prf}{\noindent\textbf{Proof.}}{\hfill $\Box$ \medskip}

\newcommand{\F}{\mathbb{F}}

\newcommand{\GL}{\textrm{GL}}
\newcommand{\SL}{\textnormal{SL}}

\newcommand{\outp}{\operatorname{Show}}

\SetKwInOut{Modified}{Modifies}
\SetKwInOut{Uses}{Uses}
\SetKwInOut{Input}{Input}

\begin{document}

\title{Showcasing straight-line programs with memory via matrix Bruhat decomposition}
\author[1]{Alice C. Niemeyer\thanks{This work was supported by the SFB-TRR 195 'Symbolic Tools in Mathematics and their Application' of the German Research Foundation (DFG).}}
\author[2]{Tomasz Popiel}
\author[3]{Cheryl E. Praeger}
\author[4]{Daniel Rademacher$^{*}$}

{ \small{
\affil[1]{Chair for Algebra and Representation Theory, RWTH Aachen University, Templergraben 55, 52062 Aachen, Germany. Email: alice.niemeyer@art.rwth-aachen.de}
\affil[2]{School of Mathematics, Monash University, Wellington Road, Clayton VIC 3800, Australia. Email: Tomasz.Popiel@monash.edu}
\affil[3]{Centre for the Mathematics of Symmetry and Computation, The University of Western Australia, 
35 Stirling Highway, Crawley 6009 WA, Australia. Email: cheryl.praeger@uwa.edu.au}
\affil[4]{Chair for Algebra and Representation Theory, RWTH Aachen University, Templergraben 55, 52062 Aachen, Germany. Email: rademacher@art.rwth-aachen.de}
}}
\maketitle

\begin{abstract}
\noindent
We  suggest  that   straight-line  programs  designed  for  algebraic
computations  should  be  accompanied  by a  comprehensive  complexity
analysis  that  takes into  account  both  the  number of  fundamental
algebraic  operations needed,  as well  as memory  requirements arising during
evaluation. We introduce an approach for formalising this idea and, as illustration,  
construct and analyse straight-line programs
for the Bruhat decomposition of $d\times d$ matrices with determinant $1$ over a
finite field of order $q$ that have length $O(d^2\log(q))$ and require
storing only $O(\log(q))$ matrices during evaluation.
\end{abstract}

\section{Introduction}
We propose  a comprehensive approach to the  analysis of straight-line
programs for use in  algebraic computations. Our approach facilitates exhaustive
complexity analyses which account  for both the number of fundamental
algebraic  operations, as  well as  memory/storage requirements
arising  during evaluation.  Our aim is to crystallise the ideas underpinning 
existing  data structures, such as those used in {\sf  GAP}~\cite{GAP} and 
{\sf  Magma}~\cite{Magma}, by introducing a data structure which we call a 
\emph{straight-line program with memory} (MSLP). We demonstrate the effectiveness 
of our approach by constructing MSLPs for the Bruhat decomposition of $d\times d$
matrices $g$ with determinant $1$ over a finite field, 
namely the decomposition $g = u_1 w u_2 $ with $u_1,u_2$ lower-unitriangular 
matrices and $w$ a monomial matrix. Our methods also yield a decomposition $g = u_1 h w' u_2$, where $h$ is a diagonal matrix and $w'$ a signed permutation matrix. In particular, we prove the following.

\begin{Thm} \label{introThm}
Let  $q=p^f$  for   some  prime  $p$  and  $f\geq   1$.  Given a matrix 
$g\in \textnormal{SL}(d,q)$, there  is  a
straight-line  program  to  compute  the  Bruhat  decomposition  of  $g$ which has length $c d^2
\log(q)$ for some  absolute constant $c$ and requires  storing at most
$2f + 18$ matrices in memory simultaneously during evaluation. Moreover, computing such a straight-line program for $g$ requires $\mathcal{O}(d^3)$ finite field operations.
\end{Thm}

Our MSLPs for this example are based on the algorithm described by Taylor \cite[p.~29]{Taylor}  and return  the  Bruhat decomposition  in
terms  of  the Leedham-Green--O'Brien~\cite{LGO} {\em  standard  generators}  of  $\SL(d,q)$, that is,  in a form that can be readily used for evaluations in {\em black box} special linear groups.
We hope  that the  concept of
MSLPs  will  lead to  more  transparent  complexity  and memory  usage
analyses in a wider context, as motivated below. \\

\noindent  {\em  Straight-line  programs  in  computational  algebra.}
Straight-line programs (SLPs) have long been used in computer science,
as programs  without branching or loops, as  discussed by B\"urgisser
et al.~\cite[Section~4.7]{BCS}.  Nowadays they are a  powerful tool in
many   areas,    including   genetics~\cite{Borgesetal},   compression
algorithms~\cite{ClaudeNavarro09},  and   in  complexity  analyses  of
algebraic computations~\cite{Lynch80}.   In our  context, an SLP  is a
sequence  of  instructions where  each  instruction  either copies  an
element  of  a given  (input)  set or  utilises  only  the results  of
previous instructions.

A seminal  paper of  Babai and Szemeredi~\cite{BabaiSzemeredi}  led to
the use of SLPs in  the analysis of algorithms for computational group
theory, and eventually to  practically efficient algorithms. 
Their paper introduced SLPs to the computational group theory community,
as well as other fundamental concepts that are now used widely in algorithms for groups, 
such as black box groups.
One of the  classical results
proved by Babai and Szemeredi~\cite{BabaiSzemeredi} is that every element
of a finite black  box group of order $n$ can be  reached by an SLP of
length at most $(1+\log(n))^2$.

The monograph by Kantor and Seress \cite{KS} led to SLPs  playing a crucial
role  in the  design and  analysis of  constructive  group recognition
algorithms;  see also  the book  by  Seress~\cite{S} or  the paper  by
O'Brien~\cite{OB} for an overview.  SLPs  now form an integral part of
the  fundamental machinery  in general  group  recognition algorithms.
These include the recognition tree of Leedham-Green and O'Brien~\cite{LGO2},
which, for example, allows the computation of the composition factors of a finite matrix group in polynomial time \cite{HLO} and is  implemented in the  computational algebra package {\sf  Magma}~\cite{Magma}, as well as a data structure
proposed by Neunh\"offer  and Seress~\cite{NS} and implemented in
{\sf  GAP}~\cite{GAP}.  In this context, the  role of SLPs in computational group theory
has  shifted.  Intended  originally as  a compact  way to  encode long
group computations to obtain a particular group element, they now need
to  allow efficient  {\em evaluation},  often in  a preimage  under a
homomorphism of the group in which they were constructed. 

One important task in this context is writing elements of classical groups as words in standard generators using SLPs. This is done in {\sf  Magma}~\cite{Magma} using the results of Elliot Costi \cite{Costi} and in GAP using the results of this paper see Section \ref{sec5}. Other rewriting algorithms also exist, for example Cohen et al.\ \cite{Cohen} present algorithms to compute with elements of finite Lie groups.\\

\noindent {\em Evaluation  of SLPs.} The evaluation of an SLP amounts
to executing  its instructions, replacing the (input)  set of elements
by a set of elements of interest.  An SLP encountered in a recognition
algorithm might record a group  computation to obtain a particular element in
the concretely represented  group and then, at a  later stage, it may need to
be evaluated  in a different  group to yield
a carefully  crafted element. Evaluation in the second group might require a  lot more
memory, and computation in this group might  be far less efficient. 
Hence,  efficient evaluation of  SLPs in
such less favourable circumstances must be addressed in practice.

A bottleneck of the original concept  of SLPs was the lack of a formal
means for identifying which subset of the evaluated instructions would
be required to evaluate subsequent instructions. Thus, when evaluating
an SLP on  a given set of group generators and  input elements, it was
hard to keep track of  how many intermediately computed elements would
no longer be  required for the rest of  the evaluation.  Consequently,
for  the complexity  analysis  of an SLP of  length
$\ell$, the  upper bound for the memory  requirement during evaluation
was  $\ell$ group elements.  
In many cases, $\ell$ is not a constant but an increasing function of the size 
of the input, and storing $\ell$ group elements may not be possible.

The implementation of SLPs in {\sf GAP} (see \texttt{straight.gi} in the
{\sf GAP} library~\cite{GAP}) addresses this issue by allowing an  
already constructed  SLP to be analysed
via the function  \textsc{SlotUsagePattern} written by Max Neunh\"offer. Information
about how to evaluate the SLP efficiently is recorded, so that no unnecessary
elements are  stored during  subsequent evaluations.  
This ensures the best possible use of memory for a given SLP in practice, but 
does not yield an upper bound for the memory requirement in a theoretical analysis.
Moreover, the result of \textsc{SlotUsagePattern} improves the memory usage but it is not necessarily optimized overall and, hence, the number of slots can still be greater than the number of slots of a carefully computed MSLP. It should also be mentioned that in some cases the number of slots can even be smaller than that of a constructed MSLP but it is not possible to predict this without a careful analysis which would result in an MSLP construction as in this paper.
A different approach is taken by B\"a\"arnhielm and Leedham-Green~\cite{BaarnhielmLeedhamGreen12}, who 
also identify the problem of storing too many intermediate elements.
They are concerned with writing
efficient SLPs to reach randomly  generated group elements, and propose
a data structure that contains an additional entry in each
instruction to record how many times this instruction will be accessed in the SLP. By not storing elements with a value of $0$ during an evaluation this ensures that no unnecessary memory is used during subsequent evaluations. \\

\noindent {\em  Straight-line programs with memory.}  While the  
approach  taken by B\"a\"arnhielm
and Leedham-Green,  and the approach in  {\sf GAP} are  tailored to  yield efficient
evaluation  of existing  SLPs,  such  as those  produced  by a  random
element  generator,  our  purpose  is  different.  
We aim to design efficient SLPs for specific tasks from the start, and to include 
memory assignments as part of the
data  structure, with  the  goal to  minimise  storage.  Our  approach
requires precise knowledge of the underlying computations for a given task to construct
an SLP that can be evaluated in a possibly different, computationally
less favourable algebraic structure, while storing only a small number
of intermediate  elements. In this  sense, our MSLPs  are custom
built.   Once an  MSLP  is constructed,  we  know how  much memory  is
required  during  evaluation,   making  the  evaluation  process  more
transparent.  This brings the construction and analysis of SLPs closer
to the aforementioned issues faced in an implementation.
We  hope  that our  approach  will  lead to  SLPs that  facilitate
efficient evaluation for many  important procedures, where most of the
intermediate  elements constructed  are quite  specific,  for example,
where particular  elements must be reached  from particular generators
of the group. We demonstrate  the concept by considering a case study,
of the Bruhat decomposition of a $d\times d$ matrix with determinant $1$ over a
finite  field  of order  $q$,  analysing  the  length and  the  memory
requirements  as functions of  $d$ and  $q$. Our example demonstrates that 
keeping track of memory requirements in the design of an SLP 
can lead to SLPs that are extremely efficient in terms of storage 
(see Theorem~\ref{introThm}).

The  paper  is structured  as  follows.   MSLPs  are  formally  defined  and  their  evaluation
discussed in  Section~\ref{sec2}, where some simple  examples are also
given.   The  in-depth Bruhat  decomposition  example is  investigated
across  Sections~\ref{sec3}   and~\ref{sec:Taylor}. In Section~\ref{sec3} we write a monomial matrix as a product of a diagonal and a signed permutation
matrix. In Section~\ref{sec:Taylor} we compute the Bruhat decomposition and draw  together   the  necessary  results  to  prove
Theorem~\ref{introThm}. 
We then comment briefly on the complexity in Section~\ref{sec6} and on a {\sf GAP} implementation of our Bruhat decomposition 
algorithm in Section~\ref{sec5}.

\section{Straight-line programs with memory} \label{sec2}

\subsection{Definitions} \label{ssec:def}

A {\em straight-line program with memory} (MSLP) is a sequence $S=[I_1,\ldots,I_n]$ together with a positive integer $b$ such that each {\em instruction} $I_r$, for $r \le n$, is formally one of the following: 
\begin{itemize}
\item[(i)] $m_k \leftarrow m_i$ with $i,k \in \{ 1,\ldots,b\}$;
\item[(ii)] $m_k \leftarrow m_i\ast m_j$ with $i,j,k \in \{ 1,\ldots,b\}$;
\item[(iii)]  $m_k \leftarrow m_i^{-1}$ with $i,k \in \{ 1,\ldots,b\}$;
\item[(iv)]  $\outp(A)$ where $A \subseteq \{1, \ldots, b \}$.
\end{itemize}
The positive integer $b$ is called the {\em memory quota} of $S$, and $S$ is said to be a {\em $b$-MSLP}. 
The positive integer $n$ is called the {\em length} of $S$, and the empty sequence is also permitted, with {\em length} $0$. 
The meaning of these instructions is revealed through {\em evaluation} of the MSLP, as follows.

The $b$-MSLP $S$ may be {\em evaluated} with respect to an ordered list $M$ 
of length $b$ of elements of a group $G$. 
The idea is that the {\em memory} $M$ will store those group elements 
that are needed in the evaluation process, with certain elements being overwritten as the
evaluation proceeds, in order to minimise the number of elements stored at any given time.
Let $M[k]$, for $1 \leq k \leq b$, denote the $k$th element, or {\em memory slot}, of $M$. 
The {\em value} of $S$ at $M$ from $s$ to $t$, where either $s=t=0$ or $1\leq s\leq t\leq n$, 
is denoted by $\operatorname{Eval}(S,M,s,t)$ and obtained as follows.

If $s=t=0$ then $\operatorname{Eval}(S,M,s,t)$
is a list of length 1 containing the identity element of $G$. 
If $1\leq s\leq t\leq n$ then for each $r\in\{s, \ldots, t\}$ in order, we perform one of the following steps:
\begin{itemize}
\item[(i)] If $I_r$ is the instruction
$m_k \leftarrow m_i$ for some $i,k\in\{1,\ldots,b\}$ then
store $M[i]$ in memory slot $M[k]$.
\item[(ii)] If $I_r$ is the instruction
$m_k \leftarrow m_i\ast m_j$ with $i,j,k \in \{ 1,\ldots,b\}$
then store in $M[k]$ the product $M[i] \cdot M[j]$ (evaluated in the group $G$).
\item[(iii)] If $I_r$ is the instruction
$m_k \leftarrow m_i^{-1}$ with $i,k \in \{ 1,\ldots,b\}$
then store in $M[k]$ the inverse $M[i]^{-1}$ of the group element $M[i]$.
\item[(iv)] If $I_r$ is the instruction
$\outp( A )$ where $A \subseteq \{1, \ldots, b \}$ then 
no action is required.
\end{itemize}
Note that when running $\operatorname{Eval}(S,M,s,t)$ for a group $G$, the computations in steps (ii) and (iii) are performed in $G$. If, for $r=t$ we perform step (i), (ii) or (iii), and hence store an element in $M[k]$, then $\operatorname{Eval}(S,M,s,t)$ is defined to be the element stored in $M[k]$. In other words, $\operatorname{Eval}(S,M,s,t)$ returns the element in the memory slot $M[k]$ overwritten by the instruction $I_t$.
On the other hand, if the instruction $I_t$ was $\outp( A )$ then $\operatorname{Eval}(S,M,s,t)$ is 
defined to be the list of elements stored in memory slots $M[i]$
for $i \in A$. 

\begin{Rem}
\textnormal{Instruction type (i) above simply copies an element already in memory to a different memory slot. These instructions can arguably be disregarded for the purpose of determining the length of an MSLP, because in a practical implementation they could be handled via relabelling. 
Therefore, for simplicity we ignore instructions of the form (i) when determining the lengths of our MSLPs in Sections~\ref{sec3} and~\ref{sec:Taylor}.}
\end{Rem}

\subsection{Basic examples} \label{ssec:ex}

To demonstrate how MSLPs function in practice, we discuss MSLPs for some fundamental group operations that arise in our more involved examples in the subsequent sections. 

\subsubsection{Commutators} \label{ex:comm}

Consider computing the commutator $[g,h] = g^{-1}h^{-1}gh$ of two group elements. 
We wish never to overwrite the input elements $g$ and $h$, which we store 
in memory slots 1 and 2, respectively.
We begin by forming the product $hg$ and storing it in a new (third) memory slot. 
The element $hg$ is then overwritten by its inverse $g^{-1}h^{-1}$. 
This element is then overwritten by $g^{-1}h^{-1}g$, which is in turn overwritten by $g^{-1}h^{-1}gh = [g,h]$. 
A total of four MSLP instructions (group multiplications or inversions) are required, and only one memory slot is needed in addition to the two memory slots used to permanently store the input elements $g,h$. 
In other words, there exists an MSLP $S$ with memory quota $b=3$ and length $4$, such that when evaluated with input containing $g$ and $h$, $S$ returns output containing $[g,h]$. 
Specifically, if we take initial/input memory $M = [g,h,1]$ then the instructions in Table~\ref{tab1} yield final/output memory with $M[3]$ the commutator $[g,h]$. 
(Unfortunately we have here a clash of notation with $[\cdot,\cdot]$ denoting both the commutator and, potentially, a list of length 2, but the meaning should be clear.)

\begin{table}[!t]
\centering
\begin{tabular}{lll}
$r$ & $I_r$ & $M$ after applying $I_r$ \\
\hline
1 & $m_3 \leftarrow m_2*m_1$ & $M=[g,h,hg]$ \\
2 & $m_3 \leftarrow m_3^{-1}$ & $M=[g,h,g^{-1}h^{-1}]$ \\
3 & $m_3 \leftarrow m_3*m_1$ & $M=[g,h,g^{-1}h^{-1}g]$ \\
4 & $m_3 \leftarrow m_3*m_2$ & $M=[g,h,g^{-1}h^{-1}gh]$ 
\end{tabular}
\caption{3-MSLP $S = [I_1, I_2, I_3, I_4]$ for the commutator of two group elements. $\operatorname{Eval}(S,M,1,4) = M[3] = g^{-1}h^{-1}gh$ when $S$ is evaluated with initial memory $M=[g,h,1]$.}
\label{tab1}
\end{table}

\begin{table}[!t]
\centering
\begin{tabular}{lll}
$r$ & $I_r$ & $M$ after applying $I_r$ \\
\hline
1 & $m_2 \leftarrow m_1*m_1$ & $[g,g^2,1,1]$ \\
2 & $m_3 \leftarrow m_1*m_2$ & $[g,g^2,g^3,1]$ \\
3 & $m_4 \leftarrow m_3$ & $[g,g^2,g^3,g^3]$ \\
4 & $m_2 \leftarrow m_2*m_2$ & $[g,g^4,g^3,g^3]$ \\
5 & $m_4 \leftarrow m_2*m_4$ & $[g,g^4,g^3,g^7]$ \\
6 & $\operatorname{Show}({3,4})$ & $[g,g^4,g^3,g^7]$
\end{tabular}
\caption{4-MSLP $S = [I_1, I_2, I_3, I_4, I_5, I_6]$ for computing the third and seventh powers of a group element. $\operatorname{Eval}(S,M,1,6) = [g^3,g^7]$ when $S$ is evaluated with initial memory $M=[g,1,1,1]$.}
\label{tab2}
\end{table}

\subsubsection{Powering via repeated squaring} \label{ex:power}

Consider powering a group element $g$ up to $g^\ell$, say, via repeated squaring. 
This can be done by using one memory slot to store the powers $g^2, g^4, g^8, \ldots, g^{\lfloor \log_2(\ell) \rfloor}$, overwriting each $g^i$ with $g^{2i}$ while another memory slot stores a running product of $g^2$-powers that is updated (if need be) upon the calculation of each $g^i$ and eventually becomes $g^\ell$. 
This procedure is completed in at most $2\lfloor \log_2(\ell) \rfloor \leq 2\log_2(\ell)$ MSLP instructions (group multiplications). 
In other words, there exists a $3$-MSLP of length at most $2\log_2(\ell)$ that when evaluated with memory containing $g$ returns output containing $g^\ell$ (one memory slot permanently stores the input $g$ while the other two slots are used for the intermediate computations). 
Table~\ref{tab2} details an explicit and slightly more complicated example, where two powers of $g$ are computed simultaneously via a 4-MSLP, thereby also demonstrating the `Show' instruction.


\section{Bruhat decomposition: step 1} \label{sec3}
In Section~\ref{sec:Taylor} we describe an MSLP for computing the Bruhat decomposition $g=u_1wu_2$ of a matrix $g \in \SL(d,q)$ where $u_1, u_2$ are lower-unitriangular matrices and $w$ is a monomial matrix.
The algorithm we use is along the lines of that given by Taylor~\cite[p.~29]{Taylor}.  
We also note the interesting account by Strang~\cite[Section~4]{Strang} of the Bruhat decomposition and its history.

The construction in Section~\ref{sec:Taylor} outputs an MSLP which evaluates to the monomial matrix $w = u_1^{-1}gu_2^{-1}$. 
The lower-unitriangular matrices $u_1$ and $u_2$ are returned as words in the Leedham-Green--O'Brien standard generators~\cite{LGO} for $\SL(d,q)$ defined in Section~\ref{ssec:strat} below. 
In the rest of this section we explain how to express an arbitrary monomial matrix $w\in\SL(d,q)$ as a word in the Leedham-Green--O'Brien standard generators, yielding an MSLP from the standard generators to $w$. 
When concatenated with the MSLP from Section~\ref{sec:Taylor}, this gives an MSLP for the complete Bruhat decomposition of $g$.

\subsection{Standard generators for $\SL(d,q)$} \label{ssec:strat}

There are several well-known generating sets for classical groups. For example, special linear groups are generated by the subset of all transvections \cite[Theorem 4.3]{Taylor} or by two well chosen matrices, such as the Steinberg generators \cite{Steinberg}. Another generating set which has become important in algorithms and applications in the last 10-15 years is the \emph{Leedham-Green and O'Brien standard generating set} in the following called the \emph{LGO generating set}. These generators are defined for all classical groups in odd characteristic in \cite{LGO} and even characteristic in \cite{LGOEven}.

The LGO generating set offers a variety of advantages. In practice it is the generating set produced by the constructive recognition algorithms from \cite{LGOEven,LGO} as implemented in MAGMA. Consequently, algorithms in the composition tree data structure, both in MAGMA and in GAP, store elements in classical groups as words in the LGO generators. Moreover, the LGO generators can be used directly to verify representations of classical groups \cite{LGOPres}.

Therefore, we decided to base the procedures we present on a set of generators very close to the LGO standard generators. Note, that the choice of the generating set has no impact on the results as it is always possible to determine an MSLP which computes the LGO standard generators given an arbritary generating set and preface an MSLP for another application by this MSLP.

We now present  the  standard generating set for  a special linear group. These are defined by Leedham-Green and O'Brien for odd characteristic~\cite{LGO} and by Dietrich, Leedham-Green, Lübeck and O'Brien for even characteristic~\cite{LGOEven}. 

Let $q=p^f$ with $p$ a prime and $f \geq 1$, and let $d$ be a positive integer. 
Fix an ordered basis for the natural module of $\SL(d,q)$ and a primitive element $\omega \in \F_q$. Let $I_m$ denote the identity element of $\SL(m,q)$ and let $\operatorname{diag}(a_1,\ldots,a_d)$ denote the diagonal matrix with $a_i$ at position $(i,i)$.
Let 
\[
s_0 = \left( \begin{array}{rr} 0&1 \\ -1&0 \end{array} \right), \, \, \,
t_0 = \left( \begin{array}{rr} 1&1 \\ 0&1 \end{array} \right) \in \F_q^{2 \times 2}, \quad
x_0 = \left( \begin{array}{rc} 0&I_3 \\ -1&0 \end{array} \right) \in \F_q^{4 \times 4}
\] 
 and define the following matrices in $\SL(d,q)$:
\begin{align*}
&\delta = \operatorname{diag}(\omega,\omega^{-1},1,\ldots,1), \quad 
s = \left( \begin{array}{cc} s_0&0 \\ 0&I_{d-2} \end{array} \right), \quad 
t = \left( \begin{array}{cc} t_0&0 \\ 0&I_{d-2} \end{array} \right), \\
& v = \left( \begin{array}{cc} 0&I_{d-2} \\ I_2&0 \end{array} \right) \text{for $d$ even or } v = 
\left( \begin{array}{cc} 0&1 \\ -I_{d-1}&0 \end{array} \right) 
\text{for $d$ odd and} \\
& x = \left( \begin{array}{cc} x_0&0 \\ 0&I_{d-4} \end{array} \right) \text{for $d$ even or } x =  I_d 
\text{ for $d$ odd}.
\end{align*}
Note that  a small variation of these {\em standard generators} for $\SL(d,q)$ are used in {\sc Magma} \cite{Magma} as well
as in algorithms to verify presentations of classical groups, see \cite{LGOPres}, where only the generator $v$ is slightly different in the two scenarios when $d$ is even.

\subsection{Transvections}
It is well-known that special linear groups contain specific elements which are called \textit{transvections}. We introduce notation of these elements in the next definition since they are needed in multiple settings. 

\begin{Def} \label{def:transvections}
Let $(e_1,\ldots,e_d)$ denote our chosen ordered basis for $V(d,q)$. 
For $i,j \in \{1,\dotsc, n\}$ with $i \neq j$, and $\alpha \in \F_q$, define $t_{ij}(\alpha) \in \SL(d,q)$ by
\[
t_{ij}(\alpha) : e_k \mapsto e_k + \alpha \Delta_{ki} e_j \quad \text{where } 
\Delta_{ki} = \begin{cases}
1 & \text{if } k=i\\
0 & \text{otherwise}.
\end{cases}
\]
That is, the matrix for $t_{ij}(\alpha)$ has ones along the main diagonal, $\alpha$ in entry $(i,j)$, and zeroes elsewhere. Note that $t_{ij}$ is an upper triangular matrix if $j > i $ and a lower triangular matrix if $j < i$.
\end{Def}

Since only one transvection, namely $t = t_{12}(1)$, is contained in the LGO standard generators, the next lemma shows how other transvections can be constructed from the LGO standard generators. Observe that throughout the entire paper group actions are performed on the right and, therefore, $\GL(d,q)$ acts on row vectors.

For the remainder of this section we continue to have $q = p^f$ with $p$ a prime and fix a primitive element $\omega \in \F_q$. Then $(\omega^0,\dotsc, \omega^{f-1})$ is an $\F_p$-basis for the $\F_p$-vector space $\F_q$ and every element $\alpha \in \F_q$ can be expressed uniquely as a polynomial $\alpha = \sum_{0 \leq \ell < f} a_\ell \omega^\ell$ in the primitive element $\omega$ and with coefficients $a_\ell \in \F_p$. Note that the elements of $\F_p$ can be identified with the integers in $\{ 0, \dotsc, p-1 \}$.

\begin{La} \label{La:trans}
Let $\alpha\in\F_q$ with $\alpha = \sum_{0 \leq \ell < f} a_\ell \omega^\ell$. Then for $i \neq j$,
\[
t_{ij}(\alpha) = \prod_{0 \leq \ell < f} t_{ij}(\omega^\ell)^{a_\ell}.
\]
The $t_{21}(\omega^\ell)$, for $0 \leq \ell < f$, are given by 
\begin{equation} \label{eq:t21}
t_{21}(\omega^\ell) = \begin{cases}
(\delta^{-\ell} v \delta^{-\ell} v^{-1}) s t^{-1} s^{-1} (\delta^{-\ell} v \delta^{-\ell} v^{-1})^{-1} 
& \textrm{if $d$ is odd} \\
(\delta^{-\ell} x^{-1} \delta^{-\ell} x) s t^{-1} s^{-1} (\delta^{-\ell} x^{-1} \delta^{-\ell} x)^{-1} 
& \textrm{if $d$ is even}.
\end{cases}
\end{equation}
The other $t_{ij}(\alpha)$, and in particular the $t_{ij}(\omega^\ell)$, are then computed recursively:
\begin{align}
\label{eq:t32}
t_{32}(\alpha) &= (xv^{-1}) t_{21}(\alpha) (xv^{-1})^{-1} \quad \mbox{if $d$ is even}, \\
\label{eq:transRec}
t_{i(i-1)}(\alpha) &= \begin{cases}
v t_{(i-1)(i-2)}(\alpha) v^{-1}
& \textrm{if $d$ is odd} \\
v^{-1} t_{(i-2)(i-3)}(\alpha) v 
& \textrm{if $d$ is even and $i\neq 3$,}
\end{cases} \\
\label{eq:comm}
t_{ij}(\alpha) &= [t_{i(j+1)}(1),t_{(j+1)j}(\alpha)] = [t_{i(i-1)}(\alpha),t_{(i-1)j}(1)] \quad \mbox{if $i-j>1$}.
\end{align}
\end{La}

\begin{Prf}
For $\beta, \gamma \in \F_q$ direct computation shows that $t_{ij}(\beta) t_{ij}(\gamma) = t_{ij}(\beta + \gamma)$ and it follows that $t_{ij}(\sum_\ell a_\ell \omega^\ell) = \prod_\ell t_{ij}(\omega^\ell)^{a_\ell}$, where each exponent $a_\ell$ is identified with an integer in $\{ 0, \dotsc ,p-1 \}$ as above.
The final assertion \eqref{eq:comm} follows from~\cite[Lemma~5.7]{Taylor}. 
Let us now justify the claimed expression \eqref{eq:t21} for $t_{21}(\omega^\ell)$. 
First suppose that $d$ is odd. Then in terms of the basis $(e_1,\dotsc,e_d)$ the action of $v$ is given by
\begin{align} \label{ActionV}
v : \begin{cases}
e_1 \mapsto e_d \\
e_i \mapsto -e_{i-1}, \quad i=2,\ldots,d
\end{cases} 
\end{align}
and hence, for any diagonal matrix $h = \operatorname{diag}(b_1,\ldots,b_d)$,
\begin{align*}
& e_1vhv^{-1} = e_dhv^{-1} = b_de_dv^{-1} = b_de_1, \\
& e_ivhv^{-1} = -e_{i-1}hv^{-1} = -b_{i-1}e_{i-1}v^{-1} = b_{i-1}e_i, \quad i=2,\ldots,d.
\end{align*}
That is, $vhv^{-1} = \operatorname{diag}(b_d,b_1,\ldots,b_{d-1})$. Since $\delta^{-\ell} = \operatorname{diag}(\omega^{-\ell},\omega^{\ell},1,\ldots,1)$ this implies that $v \delta^{-\ell} v^{-1} = \operatorname{diag}(1,\omega^{-\ell},\omega^\ell,1,\ldots,1)$. Therefore, 
$
\delta^{-\ell} v \delta^{-\ell} v^{-1} = \operatorname{diag}(\omega^{-\ell},1,\omega^\ell,1,\ldots,1),
$
and the expression for $t_{21}(\omega^\ell)$ can be verified directly by matrix multiplication, with the upper-left $3 \times 3$ block of 
\[
(\delta^{-\ell} v \delta^{-\ell} v^{-1}) s t^{-1} s^{-1} (\delta^{-\ell} v \delta^{-\ell} v^{-1})^{-1} 
\]
taking the form
\[
\left( \begin{array}{lll} \omega^{-\ell} & 0 & 0 \\ 0 & 1 & 0 \\ 0 & 0 & \omega^\ell \end{array} \right) 
\left( \begin{array}{lll} 1 & 0 & 0 \\ 1 & 1 & 0 \\ 0 & 0 & 1 \end{array} \right) 
\left( \begin{array}{lll} \omega^\ell & 0 & 0 \\ 0 & 1 & 0 \\ 0 & 0 & \omega^{-\ell} \end{array} \right) 
= 
\left( \begin{array}{lll} 1 & 0 & 0 \\ \omega^\ell & 1 & 0 \\ 0 & 0 & 1 \end{array} \right).
\]
Now suppose that $d$ is even.
The fact that $x^{-1}\delta x = \operatorname{diag}(1,\omega,\omega^{-1},1,\ldots,1)$ is verified by noting that $x$ fixes $e_5,\ldots,e_d$ and $x : (e_1,e_2,e_3,e_4) \mapsto (e_2,e_3,e_4,-e_1)$.
So in this case
\[  \delta^{-\ell} x^{-1} \delta^{-\ell} x = \operatorname{diag}(\omega^{-\ell},1,\omega^\ell,1,\ldots,1),\]
and the claimed expression for $t_{21}(\omega^\ell)$ follows as in the case of $d$ odd.

Next we check the expression \eqref{eq:transRec} for $t_{i(i-1)}$ in the case where $d$ is odd. 
We must check that $v t_{i(i-1)}(\alpha) v^{-1} = t_{(i+1)i}(\alpha)$ (for $i \neq d$), namely that the left-hand side maps $e_k$ to $e_k + \alpha \Delta_{k(i+1)} e_i$. 
Recall (\ref{ActionV}) for $d$ odd.
If $k \neq i+1$ then $e_kv \not \in \operatorname{span} \{ e_i \}$, so $t_{i(i-1)}(\alpha)$ fixes $e_kv$ and hence 
$e_k v t_{ij}(\alpha) v^{-1} = e_k v v^{-1} = e_k.$ 
For $k=i+1$ we have (remembering that by assumption here $i \neq d$)
\[
e_{i+1} v t_{i(i-1)}(\alpha) v^{-1} = -e_i t_{i(i-1)}(\alpha) v^{-1} = (-e_i - \alpha e_{i-1}) v^{-1} = e_{i+1} + \alpha e_i.
\]

Recall that $x$ fixes $e_5,\ldots,e_d$ and maps $(e_1,e_2,e_3,e_4)$ to $(e_2,e_3,e_4,-e_1)$. To verify \eqref{eq:t32}, we have $e_3(xv^{-1}) = e_4v^{-1} = e_2$ and hence
\begin{align*}
e_3 (xv^{-1}) t_{21}(\alpha) (xv^{-1})^{-1} &= e_2 t_{21}(\alpha) v x^{-1} = (e_2 + \alpha e_1) v x^{-1} \\
&= (e_4 + \alpha e_3) x^{-1} = (e_3 + \alpha e_2) = e_3 t_{32}(\alpha),
\end{align*}
while $e_k (xv^{-1}) \not \in \operatorname{span} \{ e_2 \}$ if $k\neq 3$, so in this case $e_k (xv^{-1}) t_{21}(\alpha) (xv^{-1})^{-1} = e_k (xv^{-1}) (xv^{-1})^{-1} = e_k = e_k t_{32}(\alpha)$. 

We now verify \eqref{eq:transRec} for $d$ even. 
For $i$ even (and $i\neq d$) we need to check that $v^{-1} t_{i(i-1)}(\alpha) v = t_{(i+2)(i+1)}(\alpha)$, namely that the left-hand side maps $e_{i+2}$ to $e_{i+2} + \alpha e_{i+1}$ and fixes $e_k$ when $k \neq i+2$.
Recall that, for $d$ even,
\[
v : \begin{cases}
e_j \mapsto e_{j+2}, \quad j=1,\ldots,d-2 \\
(e_{d-1},e_d) \mapsto (e_1,e_2).
\end{cases}
\]
If $k \neq i+2$ then $e_kv^{-1} \not \in \operatorname{span} \{ e_i \}$, so $t_{i(i-1)}(\alpha)$ fixes $e_kv^{-1}$ and hence $v^{-1} t_{i(i-1)}(\alpha) v$ fixes $e_k$, while
\[
e_{i+2} v^{-1} t_{i(i-1)}(\alpha) v = e_i t_{i(i-1)}(\alpha) v = (e_i + \alpha e_{i-1}) v = e_{i+2} + \alpha e_{i+1}.
\]
For $i$ odd we again use the property $v^{-1} t_{i(i-1)}(\alpha) v = t_{(i+2)(i+1)}(\alpha)$ (note that the above proof does not depend on the parity of $i$), together with equation \eqref{eq:t32}. 
\end{Prf}

For later algorithms we need to know (i) the number of matrix operations required to compute $t_{ij}(\alpha)$ for an arbitrary polynomial $\alpha = \sum_{\ell=0}^{f-1} a_\ell \omega^\ell$ and arbitrary distinct $i,j$, and (ii) the maximum number of group elements that need to be kept in memory simultaneously during this computation.

\begin{La} \label{La:arbTrans}
Let $\lambda = 2\log_2(q)+f-1$ and $b=f+3$, and let $\alpha = \sum_{\ell=0}^{f-1} a_\ell \omega^\ell$, with $\alpha_0,\ldots,\alpha_{f-1} \in \F_p$ and $i,j$ be distinct positive integers at most $d$. 
There exists a $b$-MSLP $S$ of length at most $\lambda$ such that if $S$ is evaluated with memory containing the transvections $t_{ij}(\omega^\ell)$ for $\ell=0,\ldots,f-1$, then $S$ returns memory containing the transvection  $t_{ij}(\alpha) = \prod_{\ell=0}^{f-1} t_{ij}(\omega^\ell)^{a_\ell}$. 
\end{La}

\begin{Prf}
Powering each $t_{ij}(\omega^\ell)$ up to $t_{ij}(\omega^\ell)^{a_\ell}$ via repeated squaring costs at most $2\log_2(a_\ell) \leq 2\log_2(p-1) < 2\log_2 (p)$ MSLP instructions whilst storing at most two elements (see Section~\ref{ssec:ex}(ii)). 
So computing {\em all} of the $t_{ij}(\omega^\ell)^{a_\ell}$ costs at most $2f\log_2(p) = 2\log_p(q)\log_2(p) = 2\log_2(q)$ instructions, and then an extra $f-1 = \log_p(q) - 1$ instructions are needed to obtain $t_{ij}(\alpha)$. 
Recall from Lemma \ref{La:trans} that $t_{ij}(\alpha) = \prod_{0 \leq \ell < f} t_{ij}(\omega^\ell)^{a_\ell}$. The memory quota increases by only one element, since each $t_{ij}(\omega^\ell)^{a_\ell}$ can be computed in turn, multiplied by the product of the previously computed transvections, and then discarded. 
So the total memory quota is $b=f+3$. 
\end{Prf}

\subsection{Strategy for rewriting monomial matrices} \label{ssec:actualStrat}

Let $M \leq \SL(d,q)$ be the subgroup of monomial matrices. Let $N = \langle s,x,v \rangle$ with $s,x,v$ as in Section \ref{ssec:strat}. Then $N$ is the subgroup of $\SL(d,q)$ whose matrices with respect to the standard basis $(e_1, \dotsc, e_d)$ are signed permutation matrices (namely, they have exactly one nonzero entry in each row and column and this entry is $1$ or $-1$). The aim of this section is to write a monomial matrix $w \in \SL(d,q)$ as a word in the LGO standard generators. This is achieved in two steps by writing $w$ as a product $w = hw'$ of a diagonal matrix $h$ and a signed permutation matrix $w'$. We compute words in the LGO standard generators separately for $w'$ in Section \ref{ssec:perm} and for $h$ in Section \ref{ssec:diag}. In order to find a word for a signed permutation matrix in the LGO standard generators efficiently, we employ two homomorphims $\Psi \colon M \rightarrow \operatorname{Sym}(d)$ and $\Phi \colon N \rightarrow C_2 \wr \operatorname{Sym}(d)$. Note that these homomorphisms have different domains and that $N \subseteq M$.

Define $\Psi$ to be the homomorphism from $M$ onto the group $\operatorname{Sym}(d)$ such that $w \in M$ permutes the subspaces $\langle e_1 \rangle,\ldots,\langle e_d \rangle$ in the same way as $\Psi(w)$ permutes $1,\ldots,d$. Note that for $w, w'\in M$ with $\Psi(w) = \Psi(w')$ the product $w(w')^{-1}$ is a digonal matrix. The homomorphism $\Phi$ is defined below.

Suppose first that $d$ is odd. 
Observe that $\Psi(v^{-1})$ is the $d$-cycle $(1 \ldots d)$ and $\Psi(s)$ is the transposition $(1\;2)$. 
We first construct an MSLP $S$ that, when evaluated with input memory containing the generating set $\{ (1 \ldots d),(1\;2) \}$ for $\operatorname{Sym}(d)$, outputs, as a word in these generators, a permutation $\pi_w \in \operatorname{Sym}(d)$ such that $\Psi(w)= \pi_w$. 
The MSLP $S$ is described in Section~\ref{ssec:perm} and has length $\mathcal{O}(d \operatorname{log}(d))$.
One could evaluate $S$ with input containing $\{v^{-1},s \}$ to obtain, as a word in $v$ and $s$, a signed permutation matrix $w' \in \SL(d,q)$ such that $\Psi(w')=\Psi(w)$. Therefore, we could construct the diagonal matrix $h := w (w')^{-1}$. However, evaluating $S$ on $\{v^{-1},s \}$ has complexity $\mathcal{O}(d^3 \cdot d \operatorname{log}(d)) = \mathcal{O}(d^4\operatorname{log}(d))$. We avoid evaluating $S$ in $\SL(d,q)$ by working with permutations instead. We define the following homomorphism.

\begin{Def}\label{SecondHom}
Let $w \in \SL(d,q)$ be a signed permutation matrix. Then we identify $w$ with an element $\sigma_w \in \operatorname{Sym}(2d)$ as follows:
\begin{align*}
    &i^{\sigma_w} = j \qquad \text{ and } \quad (i + d)^{\sigma_w} = j + d \, \, \text{ if } w_{ij} = 1, \text{ and } \quad \\
    &i^{\sigma_w} = j+d \, \, \text{ and } \quad (i + d)^{\sigma_w} = j \quad \quad \text{ if } w_{ij} = -1.
\end{align*}
This yields a monomorphism $\Phi \colon N \rightarrow \operatorname{Sym}(2d), w \mapsto \sigma_w$. 
\end{Def}

Given a permutation $\sigma \in \Phi(N)$, the signed permutation matrix $\Phi^{-1}(\sigma)$ is uniquely defined and can easily be written down in $\mathcal{O}(d^2)$ steps. We exploit this fact by noting that $s,x,v \in N$ whence the permutation $\tau$ we obtain by evaluating $S$ on $\{  \Phi(v^{-1}), \Phi(s) \}$ is an element in $\Phi(N)$. Clearly, $\Phi^{-1}(\tau)$ equals the matrix $w'$ we obtain by evaluating $S$ on $\{v^{-1},s \}$ as above. Moreover, $\Psi(\Phi^{-1}(\tau)) = \Psi(w)$ and, hence, $h := w\big(\Phi^{-1}(\tau)\big)^{-1}$ is the desired diagonal matrix $h$. 
The evaluation of $S$ on $\{\Phi(v^{-1}),\Phi(s) \}$ has complexity $\mathcal{O}(d \cdot d\operatorname{log}(d)) = \mathcal{O}(d^2\operatorname{log}(d))$ instead of $\mathcal{O}(d^4)$.

Finally, we construct a second MSLP, described in Section~\ref{ssec:diag}, that writes a diagonal matrix $h \in \SL(d,q)$ as a word in the standard generators of $\SL(d,q)$ (when evaluated with these generators as input). 
Combining the constructions in Sections~\ref{ssec:perm} and~\ref{ssec:diag} yields, as required, the monomial matrix
\[
w=hw'
\]
as a word in the LGO standard generators of $\SL(d,q)$. 

The only difference for the case where $d$ is even is that the standard generator $v$ does not have the property $\Psi(v^{-1}) = (1\ldots d)$; more to the point, $\Psi(v) = (1\; 3 \ldots d\!-\!1)(2 \; 4 \ldots d)$ is not a $d$-cycle in this case. 
To overcome this issue, it suffices to replace $v^{-1}$ in the above argument by a word $z$ in $v$ and $s$ such that $\Psi(z)$ is a $d$-cycle, and then replace the $d$-cycle $(1 \ldots d)$ in Section~\ref{ssec:perm} by $\Psi(z)$ via a straightforward relabelling. 
For example, one can set $z=sv$, so that $\Psi(z) = (1\; 4\; 6 \ldots d\; 2 \; 3\; 5 \; \ldots d\!-\!1)$.

\subsection{Permutations} \label{ssec:perm}

As explained above, we assume we are given a monomial matrix $w \in  M \leq \SL(d,q)$ and hence, via the homomorphism $\Psi : M \rightarrow \operatorname{Sym}(d)$, a permutation
\[
\pi_w = i \mapsto i^{\pi_w}, \quad i=1,\ldots,d.
\]
We require an MSLP that writes $\pi_w$ as a word in $\{\Phi(v^{-1}), \Phi(s)\}$. Let $\nu_1 = \Phi(v) = (1\;d\;d\!-\!1\;\ldots\;2)$ and $\sigma_1 = \Phi(s) = (1\;2)$. We solve this problem more generally by writing an arbitrary element $\pi \in \operatorname{Sym}(d)$ as a word in $\{\nu_1,\sigma_1\}$, that is
we seek an MSLP that, when evaluated with input $\nu_1$ and $\sigma_1$, outputs $\pi$ as a word in these two generators of $\operatorname{Sym}(d)$. 
Let us assume that the inverse $(1\;2\;\ldots\;d)$ of $\nu_1$ is also given as input.
We follow the strategy of the Schreier--Sims algorithm~\cite{Sims} exploiting the fact that $\{1,\dotsc,d-1\}$ is a base for $\operatorname{Sym}(d)$. For $i \in \{ 2 , \dotsc, d \}$ define 
\begin{equation} \label{SiAndVi}
\sigma_i := (i\;i+1), \quad \mbox{and } \nu_i := (i\;d\;d\!-\!1\;\ldots\;i+1).
\end{equation}
Observe that $\langle \sigma_i, \nu_i \rangle = \operatorname{Sym}(d)_{1} \cap \cdots \cap \operatorname{Sym}(d)_{i-1}$ is the pointwise stabiliser of $1,\dotsc,i-1$. Note that $\sigma_i$ can be computed recursively from $\sigma_{1}$ and $\nu_1$ via
\begin{equation} \label{recS}
\sigma_i = \nu_1 \sigma_{i-1} \nu_1^{-1}.
\end{equation}
Moreover, $\nu_i$ can be computed recursively from $\sigma_{i-1}$ and $\nu_1$ via
\begin{equation} \label{recV}
\nu_i = \sigma_{i-1} \nu_{i-1}.
\end{equation}
We show that for $\pi \in \operatorname{Sym}(d)$ we have
\begin{equation} \label{eq:pi}
\pi = \left( \prod_{i=1}^d \nu_i^{i^\pi-1} \right)^{-1},
\end{equation}
where $i^\pi-1$ is interpreted as a non-negative integer. First, $\pi_1 := \pi \nu_1^{1^\pi-1}$ fixes $1$, $\pi_2 := \pi_1\nu_2^{2^\pi-1}$ fixes $1$ and $2$, and so on, with
\[
\pi_d = \pi \prod_{i=1}^d \nu_i^{i^\pi-1} = 1.
\]
This proves (\ref{eq:pi}). The algorithm is given in pseudo-code in Algorithm \ref{alg:permutations}.

\begin{algorithm}[!t]\label{alg:permutations}
\caption{\textsc{WritePermutationAsWord}[$\pi$]}
\KwIn{a permutation $\pi \in \operatorname{Sym}(d)$\;}
\KwOut{An MSLP $S$ from $\{\nu,\sigma\}$ to $\pi$\;}
\BlankLine
$\tau = ()$\;
$\sigma = (1\;2)$\;
$\nu_1 = \nu = (1\;2\;\ldots\;d)^{-1}$\;
$\tilde{v}_{1} = \nu_1^{-1} = (1\;2\;\ldots\;d)$\;
\For{\textnormal{$i=1,\ldots,d$}}{
    $\tau = \tau \nu^{i^\pi-1}$\;
    $\nu = \sigma \nu$ \qquad \quad \, \, \, \tcp*[f]{\scriptsize{If $\sigma = \sigma_{i-1}$ and $\nu = \nu_{i-1}$, then $\nu_i = \sigma \nu$}}\;
    $\sigma = \nu_1 \sigma \nu_1^{-1} = \nu_1 \sigma \tilde{\nu}_{1} $ \tcp*[f]{\scriptsize{If $\sigma = \sigma_{i-1}$ and $v = \nu_{i-1}$, then $\sigma_i = \nu_1 \sigma \tilde{\nu}_{1}$}}\;
}
$\tau := \tau^{-1}$ \qquad \qquad \qquad \qquad \tcp*[h]{\scriptsize{Note $\tau = \pi$}}\;
Let $S$ be the MSLP of $\tau$\;
\Return $S$\;
\end{algorithm}


\begin{Pro} \label{Prop1}
Let $\lambda = 2d\log_2(d)+4d$, $b=8$ and let $\pi \in \operatorname{Sym}(d)$. 
Then Algorithm \ref{alg:permutations} yields a $b$-MSLP $S$ of length at most $\lambda$ such that, if $S$ is evaluated with memory containing $\{ \sigma_1,\nu_1,\nu_1^{-1} \}$, then $S$ outputs memory containing the permutation $\pi$.
\end{Pro}

\begin{Prf}
Computing the $\sigma_i$, for $2\leq i\leq d-1$, via the recursion~\eqref{recS} costs $2(d-2)$ MSLP instructions, and computing the $\nu_i$, for $2\leq i\leq d$, via~\eqref{recV} costs $d-1$ instructions. 
For each $1\leq i\leq d$, powering $\nu_i$ to $\nu_i^{i^\pi-1}$ requires at most 
\[
2\log_2(\pi(i)-1) 
\] 
instructions (see Section~\ref{ssec:ex}(ii)), and forming {\em all} of the $\nu_i^{i^\pi-1}$ therefore requires at most $2d\log_2(d)$ instructions. 
Computing and inverting the product $\prod_{i=1}^d \nu_i^{i^\pi-1}$ then requires a further $d$ instructions. 
So the permutation $\pi$ is computed as per~\eqref{eq:pi} in at most $2(d-2)+d-1+2d\log_2(d)+d$ MSLP instructions. As $2(d-2)+d-1+2d\log_2(d)+d < 2d\log_2(d) + 4d$ the result holds.

In addition to the three memory slots needed to store the input $\{ \sigma_1, \nu_1, \nu_1^{-1} \}$, the memory quota for the MSLP is determined as follows.  Only one slot is required to store the $\sigma_i$, because each $\sigma_i$ can overwrite $\sigma_{i-1}$ when computed via the recursion~\eqref{recS}. 
Similarly, one slot suffices to store the $\nu_i$. 
Two slots (at most) are required for the purpose of powering the $\nu_i$ up to $\nu_i^{i^\pi-1}$ via repeated squaring (see Section~\ref{ssec:ex}(i)): 
once each $\nu_i^{i^\pi-1}$ is computed, it can be multiplied by the product $\nu_1^{1^\pi-1} \cdots \nu_{i-1}^{(i-1)^\pi-1}$, which is stored in its own memory slot. 
So at most $3+1+1+2+1=8$ memory slots are needed for the MSLP.
\end{Prf}

\subsection{Diagonal matrices} \label{ssec:diag}

We now explain how to write a diagonal matrix in $\SL(d,q)$ as an MSLP in the LGO generators as $s$, $v$ and $x$. 
This is achieved by using specific upper and lower triangular transvections to avoid using a discrete logarithm oracle. Building on Lemma \ref{La:trans} we construct transvections which are upper triangular matrices.
Here, as per Section~\ref{ssec:strat}, $\omega$ denotes a primitive element of~$\F_q$. \\

The key idea is to transform the diagonal matrix with the help of row and column operations into the identity matrix in a way similar to an algorithm to compute the elementary divisors of an integer matrix, as described for example in \cite[Chapter 7, Section 3]{Hartley_Hawkes_Ring_Modules}. Note that row and column operations are effected by left- and right multiplications by transvections. Thus recording the row and and column operations required to transform a diagonal matrix into the identity, allows us to write the input matrix as a product of transvections. \\

We require specific upper triangular transvections which are constructed in the following lemma. Recall the definition of the transvections $t_{ij}(\alpha)$ with respect to the standard basis $(e_1,\dotsc, e_d)$ of $\F_q^d$ from Definition \ref{def:transvections}.

\begin{La} \label{La:trans2}
For all $d \in \mathbb{N}$ and $\alpha \in \F_q^*$
\begin{align*}
   s^{-1} t_{21}(- \alpha) s &= t_{12}(\alpha).
\end{align*}
Let $s_1 := x^{-1} s x$ if $d$ is even. Then
\begin{align*}
   s_1^{-1} t_{32}(- \alpha) s_1 &= t_{23}(\alpha).
\end{align*}
\end{La}

\begin{Prf}
Clearly, for $k > 2$, $e_k s^{-1} t_{21}(- \alpha) s = e_k$ as $s$ and $t_{21}(- \alpha)$ fix $e_k$. Moreover,
\begin{center}
    $e_1 s^{-1} t_{21}(- \alpha) s = (-1) e_2 t_{21}(- \alpha) s = (-1) (e_2 - \alpha e_1) s =   \alpha e_1 s - e_2 s =  \alpha e_2 + e_1$ 
\end{center}
and 
\begin{center}
    $e_2 s^{-1} t_{21}(- \alpha) s = e_1 t_{21}(- \alpha) s = e_1 s = e_2$.
\end{center}
Hence, $s^{-1} t_{21}(- \alpha) s = t_{12}(\alpha)$. Now let $d$ be even. Then 
\begin{align*}
    e_1 s_1 &= e_1 x^{-1} s x = - e_4 s x = -e_4 x = e_1, \\
    e_2 s_1 &= e_2 x^{-1} s x = e_1 s x = e_2 x = e_3, \\
    e_3 s_1 &= e_3 x^{-1} s x = e_2 s x = - e_1 x = - e_2, \\
    e_4 s_1 &= e_4 x^{-1} s x = e_3 s x = e_3 x = e_4, 
\end{align*}
and the last equality follows as for $s^{-1} t_{21}(- \alpha) s$.
\end{Prf}

As for the Bruhat decomposition, the transvections $t_{32}(\alpha)$ are only required when $d$ is even. Before proving upper bounds, we give an example of the idea.

\begin{Ex}
We consider the following diagonal matrix $h \in \SL(3,7)$:
\begin{center}
    $h := \left( \begin{array}{rrr} 3&0&0 \\ 0&2&0 \\ 0&0&6 \end{array} \right)$.
\end{center}
We show that $h$ can be transformed into the identity matrix using only row and column operations, where $\xrightarrow{\cdot^R t}$ denotes a right multiplication with the matrix $t$ and $\xrightarrow{\cdot^L t}$ denotes a left multiplication with the matrix $t$. We start by adding the second column to the first column:
\begin{center}
     $\left( \begin{array}{rrr} 3&0&0 \\ 0&2&0 \\ 0&0&6 \end{array} \right) \xrightarrow{\cdot^R t_{21}(1)}
     \left( \begin{array}{rrr} 3&0&0 \\ 2&2&0 \\ 0&0&6 \end{array} \right)$.
\end{center}
Then we subtract the second row from the first row such that the entry at position $(1,1)$ becomes 1:
\begin{center}
     $\left( \begin{array}{rrr} 3&0&0 \\ 2&2&0 \\ 0&0&6 \end{array} \right)\xrightarrow{\cdot^L t_{12}(6)}
     \left( \begin{array}{rrr} 1&5&0 \\ 2&2&0 \\ 0&0&6 \end{array} \right)$.
\end{center}
Now we add the first column to the second column and subsequently subtract the first row from the second row to re-establish a diagonal matrix:
\begin{center}
     $\left( \begin{array}{rrr} 1&5&0 \\ 2&2&0 \\ 0&0&6 \end{array} \right)\xrightarrow{\cdot^R t_{12}(2)}
     \left( \begin{array}{rrr} 1&0&0 \\ 2&6&0 \\ 0&0&6 \end{array} \right) \xrightarrow{\cdot^L t_{21}(5)}
     \left( \begin{array}{rrr} 1&0&0 \\ 0&6&0 \\ 0&0&6 \end{array} \right)$.
\end{center}
We repeat this procedure for the entry at position $(2,2)$: 
\begin{center}
     $\left( \begin{array}{rrr} 1&0&0 \\ 0&6&0 \\ 0&0&6 \end{array} \right) \xrightarrow{\cdot^R t_{32}(1)} \left( \begin{array}{rrr} 1&0&0 \\ 0&6&0 \\ 0&6&6 \end{array} \right) \xrightarrow{\cdot^L t_{23}(5)}
     \left( \begin{array}{rrr} 1&0&0 \\ 0&1&2 \\ 0&6&6 \end{array} \right)$,
\end{center}
\begin{center}
    $  \left( \begin{array}{rrr} 1&0&0 \\ 0&1&2 \\ 0&6&6 \end{array} \right)
    \xrightarrow{\cdot^R t_{23}(5)}
    \left( \begin{array}{rrr} 1&0&0 \\ 0&1&0 \\ 0&6&1 \end{array} \right)
    \xrightarrow{\cdot^L t_{32}(1)}
    \left( \begin{array}{rrr} 1&0&0 \\ 0&1&0 \\ 0&0&1 \end{array} \right)$.
\end{center}
Note that the entry at position $(3,3)$ automatically becomes $1$ as the matrix is contained in $\SL(3,7)$. Thus
\begin{center}
    $h = (t_{32}(1)t_{23}(5)t_{21}(5)t_{12}(6))^{-1} (t_{21}(1)t_{12}(2)t_{32}(1)t_{23}(5))^{-1}$.
\end{center}
\end{Ex}

The algorithm is also given in pseudo-code in Algorithm \ref{alg:diagonalMatrices}. 

\begin{algorithm}\label{alg:diagonalMatrices}
\caption{\textsc{WriteDiagonalMatrixAsWord}[$h$]}
\tcp{$X := \{ s,s^{-1},t,t^{-1},\delta,\delta^{-1},v,v^{-1},x,x^{-1} \}$, LGO gens and inverses}
\KwIn{a diagonal matrix $h \in \SL(d,q)$\; An MSLP $S$ evaluating from $X$ to the set defined in Lemma \ref{La:Preprocess};}
\KwOut{An MSLP $S'$ evaluating from $X$ to $h$\;}
\SetAlgoLined
$h_{\ell} := I_d$\;
$h_r := I_d$\;
\For{\textnormal{$i=1,\ldots,d-1$}}{
    $\alpha = h[i,i]$ \;
    \If{$\alpha \neq 1$}{
        $\beta = h[i+1,i+1]$\;
        Add column $i+1$ to column $i$ of $h$\;
        $h_r := h_r t_{i+1,i}(1)$\;
        Add $(1-\alpha)/\beta$ times row $i+1$ to row $i$ of $h$\;
        $h_{\ell} := t_{i,i+1}((1-\alpha)/\beta) h_{\ell}$\;
    }
        Add $\alpha-1$ times column $i$ to column $i+1$ of $h$\;
        $h_r := h_r t_{i,i+1}(\alpha-1)$\;
        Add $-1 \cdot\beta$ times row $i$ to row $i+1$ of $h$\;
        $h_{\ell} := t_{i+1,i}(-\beta) h_{\ell}$\;
}
$S := $ Composition of $S$ and the MSLP of $h_{\ell}^{-1}h_r^{-1}$ \;
\Return $S$\;
\end{algorithm} 

We divide the algorithm of writing MSLPs for diagonal matrices into a preprocessing and a computation phase. 

First we describe the preprocessing phase during which we initialize the memory of the MSLP to encode particular matrices which will be useful for expressing diagonal matrices as words independently of the given diagonal matrix. The constructed matrices can be reused for all diagonal matrices, and so further diagonal matrices can be expressed as MSLPs without performing the preprocessing step multiple times.

The following lemma shows how to compute the matrices of the preprocessing step. Recall that $\omega$ is a primitive element of $\F_q = \F_{p^f}$.

\begin{La} \label{La:Preprocess}
Let $\lambda = 10f + 2$ and $b=10 + 2f$ if $d$ is odd, and $\lambda = 16f + 7$ and \linebreak $b=10 + 4f$ if $d$ is even. Moreover, let $X = \{ s,s^{-1},t,t^{-1},\delta,\delta^{-1},v,v^{-1},x,x^{-1} \}$.
There exists a $b$-MSLP $S$ of length at most $\lambda$ such that if $S$ is evaluated with memory containing the set $X$ then $S$ outputs final memory containing $t_{21}(\omega^\ell)$ and $t_{12}(\omega^\ell)$ for $0 \leq \ell \leq f-1$ and additionally, when $d$ is even, $t_{32}(\omega^\ell)$ and $t_{23}(\omega^\ell)$ for $0 \leq \ell \leq f-1$ 
\end{La}

\begin{Prf}
Let $S$ denote the $b$-MSLP constructed. Let $p[i,j,\ell]$ denote the position in the memory, where $t_{ij}(\omega^\ell)$ will be output when the MSLP $S$ is evaluated, where $0 \leq \ell \leq f-1$ and $(i,j) \in \{(1,2),(2,1)\}$ if $d$ is odd and $(i,j) \in \{(1,2),(2,1), (2,3), (3,2) \}$ if $d$ is even.

Let $d$ be odd. The memory is initialized as
\begin{center}
$[s,s^{-1},t,t^{-1},\delta,\delta^{-1},v,v^{-1},x,x^{-1},1_G,\dotsc,1_G]$
\end{center}
and will be transformed as follows after applying the MSLP $S$:
\begin{center}
    $[s,s^{-1},t,t^{-1},\delta,\delta^{-1},v,v^{-1},x,x^{-1},t_{21}(\omega^0),t_{21}(\omega^1),\dotsc,t_{21}(\omega^{f-1}),\dotsc \newline
    \dotsc, t_{12}(\omega^0),t_{12}(\omega^1),\dotsc,t_{12}(\omega^{f-1})].
    $
\end{center}
We start by computing lower triangular matrices $t_{21}(\omega^\ell)$ for $0 \leq \ell \leq f-1$ (note that one could also reuse the matrices from the Bruhat decomposition computation if that algorithm was used beforehand). During the process we may store additional matrices in the slots $p[1,2,\ell]$ reserved for upper triangular matrices $t_{12}(\omega^\ell)$ for $0 \leq \ell \leq f-1$. If $f = 1$, then we do not need the additional slot $p[1,2,1]$ for the upper triangular matrix $t_{12}(1)$. If $f > 1$, then we use the two slots $p[1,2,1]$ and $p[1,2,2]$ reserved for the upper triangular matrices $t_{12}(1)$ and $t_{12}(\omega)$. In the slot $p[1,2,1]$ of $t_{12}(1) = t_{12}(\omega^0)$ we store $\delta^{-\ell}$ and in the slot $p[1,2,2]$ of $t_{12}(\omega^1)$ we store $st^{-1}s^{-1}$. 

The computation of $\delta^{1}$ requires zero operations and the computation of $st^{-1}s^{-1}$ requires two operations. During the iteration we compute $\delta^{-\ell} =$ \linebreak $\delta^{-1} \cdot \delta^{1-\ell}$ with one operation in the slot $p[1,2,1]$ of $t_{12}(\omega^0)$ and $\delta^{-\ell} v \delta^{-\ell} v^{-1} $ in three operations in the slot $p[2,1,\ell]$ of $t_{21}(\omega^\ell)$. Then it only takes three additional operations to compute $t_{21}(\omega^\ell) = (\delta^{-\ell} v \delta^{-\ell} v^{-1}) s t^{-1} s^{-1} (\delta^{-\ell} v \delta^{-\ell} v^{-1})^{-1} $. Overall one iteration needs seven operations. Since we have to do this $f$ times, this process takes $7f + 2$ operations. 

Now we compute upper triangular matrices. For this we can use lower triangular matrices and we need three operations per upper triangular matrix using Lemma \ref{La:trans2}. This requires at most $10f + 2$ operations. 

Now let $d$ be even. The same results for the transvections $t_{21}(\omega^\ell)$ and $t_{12}(\omega^\ell)$ as for $d$ odd can be obtained by replacing $v$ by $x$ in the formula for $t_{21}(\omega^\ell)$. It remains to compute $t_{32}(\omega^\ell)$ and $t_{23}(\omega^\ell)$ which can be done using Lemmas \ref{La:trans} and \ref{La:trans2}. First, we compute $xv^{-1}$ and store it in the slot $p[2,3,1]$ for $t_{23}(\omega^0)$ which takes one operation. Then we compute $t_{32}(\omega^\ell) = (xv^{-1})t_{21}(\omega^\ell)(xv^{-1})^{-1}$ for $0 \leq \ell < f$ which needs three operations per transvection, and hence $3f$ operations overall. Lastly we compute $s_1 = v s v^{-1}$ and store it in slot $p[2,3,f-1]$ which needs two operations and $t_{23}(\omega^\ell)$ for $0 \leq \ell < f$ which needs $3f$ operations overall. This requires at most $16f + 7$ operations.
\end{Prf}

We now compute upper bounds for the length and memory quota of an MSLP for expressing an arbitrary diagonal matrix $h\in\SL(d,q)$ as a word in the LGO generators, i.e.\ the computation phase of the algorithm.

\begin{Pro} \label{Prop2}
Let $\lambda = (d-1)(6\log_2(q)+7f+1)$ and $b=14 + 2f$ if $d$ is odd and $b=14 + 4f$ if $d$ is even. 
Let $h\in\SL(d,q)$ be a diagonal matrix. 
Then Algorithm \ref{alg:diagonalMatrices} yields a $b$-MSLP $S$ of length at most $\lambda$ such that if $S$ is evaluated with memory containing the LGO generators and the matrices of Lemma \ref{La:Preprocess}, then $S$ outputs final memory containing $h$.
\end{Pro}

\begin{Prf}
Algorithm \ref{alg:diagonalMatrices} computes a $b$-MSLP with the required properties. We first analyse its complexity and then prove the assertions on $b$. 

Let $i \in \{1,\dotsc,d-1 \}$. Getting the diagonal entry of $h$ at position $(i,i)$ to $1$ requires the following number of operations. We start by adding the column $i+1$ to column $i$ as in Line 5. We already have the transvection $t_{i (i+1)}(1)$ and, hence, this requires 1 operation. Adding some scalar times row $i+1$ to row $i$ to obtain a $1$ at position $(i,i)$ as in Line 6 requires at most $2\log_2(q)+f-1$ operations to compute the corresponding transvection using Lemma \ref{La:arbTrans} and 1 operation for multiplying with the output matrix. Clearing the two non-diagonal entries at positions $(i,i+1)$ and $(i+1,i)$ by addition with the $i$-th row and column requires at most $2 \cdot (2\log_2(q)+f)$ operations with the same argument as in Line 7 and 8. This yields $3 \cdot (2\log_2(q)+f) + 1$ operations. Shifting the transvections $t_{i(i+1)}(\omega^\ell)$ and $t_{(i+1)i}(\omega^\ell)$ using Lemma \ref{La:trans} requires 2 operations per transvection and, therefore, $2 \cdot 2f = 4f$ operations in total. Hence, we need $3 \cdot (2\log_2(q)+f) + 1 + 4f$ operations for one diagonal entry and for shifting the transvections. Since, we have to do this $d-1$ times, the overall number of operations is $(d-1)(3 \cdot (2\log_2(q)+f) + 1 + 4f)$.
\newline
\newline
In addition to storing the input, the memory quota for the MSLP is as follows. 
At most two memory slots are needed for repeated squaring, one slot is needed for storing a transvection and one slot is needed for the output diagonal matrix. Therefore, the MSLP requires $4$ memory slots in addition to the input.
\end{Prf}

\begin{Rem}
\textnormal{
Note that the procedure does not require any applications of a discrete logarithm oracle as was required in an earlier version of this procedure in \cite{MSLP2013}.
}
\end{Rem}

\section{Bruhat decomposition: step 2} \label{sec:Taylor}

We recall from Taylor~\cite[p.~29]{Taylor} an algorithm for obtaining the Bruhat decomposition of a matrix
\[
g = (g_{ij}) \in \SL(d,q),
\]
namely for writing $g$ in the form
\begin{equation} \label{eq:Bruhat}
g = u_1 w u_2 \quad
\text{with $u_1$, $u_2$ lower unitriangular and $w$ monomial.}
\end{equation}
The algorithm reduces $g$ to the monomial matrix $w$ by multiplying $g$ by certain transvections, namely the $t_{ij}({\alpha})$ of Definition \ref{def:transvections}.

Multiplying $g$ on the left by the transvection $t_{ij}(\alpha)$ effects an elementary row operation that adds $\alpha$ times the $j$th row to the $i$th row. 
Similarly, right multiplication by $t_{ij}(\alpha)$ effects an elementary column operation, adding $\alpha$ times column $i$ to column $j$. 
Using the {\em row} operations, one can reduce $g$ to a matrix with exactly one nonzero entry in its $d$th column, say in row $r$. 
Then the elementary {\em column} operations can be used to reduce the other entries in row $r$ to zero. 
Continuing recursively, $g$ can be reduced to a matrix with exactly one nonzero entry in each row and each column. 
Moreover, at the end of the procedure, the products of the transvections $t_{ij}(\alpha)$ on the left- and right-hand sides of $g$ both form lower unitriangular matrices, because $j<i$.
In other words, one obtains a decomposition of the form \eqref{eq:Bruhat}.
Note that the described procedure differs from the one in Taylor's book in that we multiply $g$ on the left and right by {\em lower}-triangular transvections ($j<i$), as opposed to upper-triangular transvections $(i<j)$. This allows us to achieve our objective of restricting the memory requirements.

Our aim is to determine the length and memory quota for an MSLP for the Bruhat decomposition of an arbitrary matrix $g \in \SL(d,q)$ via the above method, with the matrices $u_1$, $u_2$, $w$ returned as words in the LGO generators $s,t,v,\delta,x$ of $\SL(d,q)$ given in Section \ref{ssec:strat}.

As we run through the algorithm described by Taylor, we deal with the columns of $g$ in reverse order, beginning with column $d$. 
Suppose we have reached column $c$, for some $c \in \{ 1,\ldots,d \}$. 
At this stage, $g$ has been reduced to a matrix in which columns $c-1,\ldots,d$ have exactly one nonzero entry (and these entries are in different rows). 
This has been achieved by multiplying $g$ on the left and right by certain transvections, the products of which are lower-unitriangular matrices.

We identify the first row in which the $c$th column contains a nonzero entry, say the $r$th row; that is, we set
\[
r = r(c) := \min_{1\leq i\leq d} \{ g_{ic} \neq 0 \}.
\]
The idea is to eliminate all other entries in the $c$th column, namely to apply elementary row operations to make the entries in rows $i=r+1,\ldots,d$ of column $c$ equal to zero. Specifically, $g$ is multiplied on the left by the transvections
\begin{equation} \label{trans1}
t_{ir}(-g_{ic}g_{rc}^{-1}),
\end{equation}
which subtract $g_{ic}g_{rc}^{-1}$ times the $r$th row of $g$ from the $i$th row. 
Having `cleared' column $c$, we clear the entries in position $j=1,\ldots,c-1$ in the $r$th row by multiplying $g$ on the right by the transvections
\begin{equation} \label{trans2}
t_{cj}(-g_{rj}g_{rc}^{-1}), 
\end{equation}
which subtracts $g_{rj}g_{rc}^{-1}$ times the $c$th column of $g$ from the $j$th column. 

For the purposes of determining the cost of Taylor's algorithm in terms of matrix operations, namely determining the length of an MSLP for the algorithm, we assume that the field elements $-g_{ic}g_{rc}^{-1}$ in \eqref{trans1} (and similarly in \eqref{trans2}) are given to us as polynomials of degree at most $f-1$ in the primitive element $\omega$, where $q=p^f$ for some prime $p$. 

\subsection{Pseudocode and analysis for $d$ odd}

Pseudocode for Taylor's algorithm for obtaining the Bruhat decomposition of an arbitrary matrix $g \in \SL(d,q)$ is presented in Algorithm~\ref{alg:TaylorFull} for the case where $d$ is odd. 
The case where $d$ is even is very similar, but requires a few changes that would complicate the pseudocode. 
So, for the clarity of our exposition, we analyse the case $d$ odd here and then explain the differences for the case $d$ even in the next subsection. 
To aid the exposition and analysis, Algorithm~\ref{alg:TaylorFull} refers to several subroutines, namely Algorithms~\ref{alg:FirstTransvections}--\ref{alg:BackshiftTransvections}. In an implementation the code for the Algorithms~\ref{alg:FirstTransvections}--\ref{alg:BackshiftTransvections} would be inserted into Algorithm~\ref{alg:TaylorFull} in the lines where they are called. We present them as subroutines here to improve the readability of Algorithm~\ref{alg:TaylorFull}. However, we assume Algorithms~\ref{alg:FirstTransvections}--\ref{alg:BackshiftTransvections} have access to the variables of Algorithm~\ref{alg:TaylorFull} in an implementation and that the Algorithms~\ref{alg:FirstTransvections}--\ref{alg:BackshiftTransvections} can also use and modify variables of Algorithm~\ref{alg:TaylorFull}. The variables of Algorithm~\ref{alg:TaylorFull} modified and used by Algorithms~\ref{alg:FirstTransvections}--\ref{alg:BackshiftTransvections} are highlighted under the headings ``used'' and ``modified''.

We determine upper bounds for the length and memory quota of an MSLP for Algorithm~\ref{alg:TaylorFull}. 
Recall that we use the LGO generators $\delta,s,t,v,x$ of $\SL(d,q)$; although $x=1$ when $d$ is odd, we include $x$ here for consistency with the $d$ even case in the next subsection. 
Define the (ordered) list 
\begin{equation} \label{eq:gens}
Y = Y(w,u_1,u_2) := [ s,s^{-1},t,t^{-1},\delta,\delta^{-1},v,v^{-1},x,x^{-1},w,u_1,u_2 ].
\end{equation}
The idea is that our MSLP, when evaluated with initial memory containing $Y(g,1,1)$, should output final memory containing $Y(w,u_1,u_2)$ with $g=u_1wu_2$ the Bruhat decomposition of $g$. 
In other words, our algorithm initialises $w:=g$, $u_1:=1$ and $u_2:=1$ and multiplies $w$, $u_1$ and $u_2$ by the transvections necessary to render $g=u_1wu_2$ with $w$ monomial and $u_1,u_2$ lower unitriangular. 

As for the simpler examples considered in the previous section, here to keep the presentation clear we do not write down explicit MSLP {\em instructions}, but instead determine the cost of Algorithm~\ref{alg:TaylorFull} while keeping track of the number of elements that an MSLP for this algorithm would need to keep in memory at any given time. 
We prove the following.

\begin{Pro} \label{dOddProp}
Let $q=p^f$ with $p$ prime and $f \geq 1$, let $d$ be an odd integer with $d \geq 3$, and set $b=f+18$ and 
\[
\lambda =d^2(2\log_2(q) + 5f + 10) + 4d(\log_2(q) + 1) + (5f+1).
\]
There exists a $b$-MSLP, $S$, of length at most $\lambda$ such that if $S$ is evaluated with memory containing the list $Y(g,1,1)$ as in~\eqref{eq:gens} then $S$ outputs memory containing $Y(w,u_1,u_2)$ with $g=u_1wu_2$ the Bruhat decomposition of~$g$.
\end{Pro}

Let us also write
\[
T_i := \{ t_{i(i-1)}(\omega^\ell) \mid \ell=0,\ldots,f-1 \} \quad \text{for } i=2,\ldots,d.
\]
As noted in the previous subsection, these sets of transvections are needed to construct the transvections by which $g$ is to be multiplied.

In practice, the MSLP should be constructed in such a way that the `input' of each of the subroutines (Algorithms~\ref{alg:FirstTransvections}--\ref{alg:BackshiftTransvections}) is stored in memory when the subroutine is called and the `output' is kept in memory for the subsequent stage of Algorithm~\ref{alg:TaylorFull}. 
The cost of the subroutines is determined with this in mind; that is, for each subroutine we determine the maximum length and memory requirement for an MSLP that returns the required output when evaluated with an initial memory containing the appropriate input. 

\begin{algorithm}[!t]\label{alg:TaylorFull}
\caption{\textsc{BruhatDecompositionOdd}[$g$]}
\KwIn{the list $Y$ in \eqref{eq:gens} with $w:=g$ and $u_1,u_2 := 1$\;}
\KwOut{a monomial matrix $w\in\SL(d,q)$ and two lower unitriangular matrices $u_1,u_2\in \SL(d,q)$ such that $g=u_1wu_2$\;}
\BlankLine
compute $T_2$ from the standard generators\;
\For{\textnormal{$c=d,\ldots,1$}}{
\If{$r:= \min\{i \mid h_{ic}\} \leq d-1$}{
call subroutine \textsc{FirstTransvections}[$r$] (Algorithm~\ref{alg:FirstTransvections})\;
\If{$r \leq d-2$}{
\For{$i=r+2,\ldots,d$}{
call subroutine \textsc{LeftUpdate}[$i$] (Algorithm~\ref{alg:ShiftTransvections})\;
}
}
}
\If{$c\geq 2$}{
call subroutine \textsc{LastTransvections}[$c$] (Algorithm~\ref{alg:LastTransvections})\;
\If{$c\geq 3$}{
\For{\textnormal{$j=c-2,\ldots,1$}}{
call subroutine \textsc{RightUpdate}[$j$] (Algorithm~\ref{alg:BackshiftTransvections})\;
}
}
}
}
$u_1 := u_1^{-1}$\;
$u_2 := u_2^{-1}$\;
\Return $w$, $u_1$, $u_2$\;
\end{algorithm}

\subsubsection{Computing $T_2$ from the standard generators}

The first step of the algorithm is the one-off computation of $T_2$ from the LGO standard generators of $\SL(d,q)$. The length and memory requirement of an MSLP for this step is as follows.

\begin{La} \label{la1}
Let $\lambda=5f-1$ and $b=f+14$. 
There exists a $b$-MSLP, $S$, of length $\lambda$ such that if $S$ is evaluated with memory containing the list $Y(w,u_1,u_2)$ as in \eqref{eq:gens} then $S$ outputs memory containing $T_2$ and $Y(w,u_1,u_2)$.
\end{La}

\begin{Prf}
Recall the expression \eqref{eq:t21} for the $t_{21}(\omega^\ell)$ in the $d$ odd case, namely
\[
t_{21}(\omega^\ell) = (\delta^{-\ell} v \delta^{-\ell} v^{-1}) s t^{-1} s^{-1} (\delta^{-\ell} v \delta^{-\ell} v^{-1})^{-1}.
\]
Computing $t_{21}(1) = st^{-1}s^{-1}$ costs two matrix multiplications of elements from $Y$, and thus adds two instructions to our MSLP. 
One new memory slot, in addition to the $|Y|=13$ initial slots, is needed to store $t_{21}(1)$.
If $f\geq 2$ then we continue by observing that, for $\ell = 1,\ldots,f-1$,
\[
t_{21}(\omega^\ell) = z_\ell \hat{t} z_\ell^{-1},
\]
where $\hat{t} := v^{-1}t_{21}(1)v^{-1}$ and the $z_\ell$ are given recursively by
\[
z_\ell = \delta^{-1} z_{\ell-1} \delta^{-1} \quad \text{with } z_1 = \delta^{-1} v \delta^{-1}.
\]
Computing $\hat{t}$ requires two MSLP instructions (matrix multiplications) and one new memory slot. 
Computing the $z_\ell$ requires $2(f-1)$ instructions, but also only one new memory slot because each $z_\ell$ can overwrite $z_{\ell-1}$ as the algorithm proceeds recursively. 
Similarly, computing the $z_\ell^{-1}$ takes $(f-1)$ instructions but only one new memory slot.
Forming the $t_{21}(\omega^\ell) = z_\ell \hat{t} z_\ell^{-1}$ costs $2(f-1)$ instructions, and $f-2$ new memory slots because each $t_{21}(\omega^\ell)$ needs to be returned by the algorithm but $t_{21}(\omega^{f-1})$ can overwrite $\hat{t}$. 
In total, we require
\[
2 + 2 + 2(f-1) + (f-1) + 2(f-1) = 5f-1
\]
instructions, and
\[
[ 3 + (f-2) ] + |Y| = f+14
\]
memory slots. 
\end{Prf}

\subsubsection{Calls to Algorithm~\ref{alg:FirstTransvections}}

\begin{algorithm}[!t] \label{alg:FirstTransvections}
\caption{\textsc{FirstTransvections}[$r$]}
\Input{$r \in \mathbb{N}$\;}
\Uses{$T_2$ and $Y(w,u_1,u_2)$\;}
\Modified{$T_{r+1}$ and $Y(t_{(r+1)r}(-g_{(r+1)c}g_{rc}^{-1}) w,t_{(r+1)r}(-g_{(r+1)c}g_{rc}^{-1}) u_1,u_2)$\;}
\BlankLine
\If{$r\geq 2$}{
\For{$\ell=0,\ldots,f-1$}{
\For{$i=3,\ldots,r+1$}{
compute $t_{i(i-1)}(\omega^\ell) = v t_{(i-1)(i-2)}(\omega^\ell) v^{-1}$\;
}
}
}
compute $t_{(r+1)r}(-g_{(r+1)c}g_{rc}^{-1})$\;
$w := t_{(r+1)r}(-g_{(r+1)c}g_{rc}^{-1}) w$\; 
$u_1 := t_{(r+1)r}(-g_{(r+1)c}g_{rc}^{-1}) u_1$\;
\end{algorithm}

Having computed the $T_2$, we begin the main `for' loop of Algorithm~\ref{alg:TaylorFull}, running through the columns of $g$ in reverse order. 
Observe that $r$ takes each value $1,\dots,d$ exactly once as we run through the columns of $g$, because we are reducing the nonsingular matrix $g$ to a monomial matrix. 
If we are in the (unique) column where $r=d$ then there is no `column clearing' to do and we skip straight to the row clearing stage. 
For each other column, we start by calling the subroutine \textsc{FirstTransvections[$r$]} (Algorithm~\ref{alg:FirstTransvections}).
The role of this subroutine is to multiply the matrix $g$ on the left by the transvection
\[
t_{(r+1)r}(-g_{(r+1)c}g_{rc}^{-1}),
\] 
thereby making the $(r+1,c)$-entry of $g$ zero. 
If $r\geq 2$, it is necessary to first compute the $t_{(r+1)r}(\omega^\ell)$ (if $r=1$ then these are already stored in memory). 
The cost of an MSLP for calls to \textsc{FirstTransvections[$r$]} is as follows.

\begin{La} \label{la2}
Let $\lambda=f(2r-1) + 2\log_2(q) + 2$ and $b=f+17$. 
There exists a $b$-MSLP, $S$, of length at most $\lambda$ such that if $S$ is evaluated with memory containing the input of Algorithm~$\ref{alg:FirstTransvections}$ then $S$ returns memory containing the output of Algorithm~$\ref{alg:FirstTransvections}$.
\end{La}

\begin{Prf}
The nested `for' loop in \textsc{FirstTransvections[$r$]} computes the set of transvections $T_{r+1}$ in $2f(r-1)$ instructions (matrix operations). 
It does not increase the size 
\[
|T_2| + |Y(w,u_1,u_2)| = f+13
\] 
of the input memory, because the recursive formula for the $T_i$ allows each $T_i$ to overwrite $T_{i-1}$. 
If we assume that $-g_{(r+1)c}g_{rc}^{-1}$ is known as a polynomial of degree (at most) $f-1$ in the primitive element $\omega \in \F_p$, then we can construct the transvection $t_{(r+1)r}(-g_{(r+1)c}g_{rc}^{-1})$ from $T_{r+1}$ in at most $2\log_2(q)+f$ instructions, according to Lemma~\ref{La:arbTrans}. 
A further two instructions are then required to multiply $w$ and $u_1$ by this transvection to return the required final memory. 
In addition to the existing $f+13$ memory slots already in use, a further three slots are required while computing $t_{(r+1)r}(-g_{(r+1)c}g_{rc}^{-1})$, according to Lemma~\ref{La:arbTrans} (because the transvections to be powered up and multiplied together are already in memory), and then one more slot is needed to store this transvection prior to multiplying it by $w$ and $u_1$. 
So the MSLP requires at most
\[
2f(r-1) + (2\log_2(q)+f) + 2 = f(2r-1) + 2\log_2(q) + 2
\]
instructions and at most $(f+13)+4 = f+17$ memory slots.
\end{Prf}

\subsubsection{Calls to Algorithm~\ref{alg:ShiftTransvections}}

\begin{algorithm}[!t] \label{alg:ShiftTransvections}
\caption{\textsc{LeftUpdate}[$i$]}
\Input{$i \in \mathbb{N}$\;}
\Uses{$T_{i-1} \cup \{ t_{(i-1)r}(1) \} \cup Y(w,u_1,u_2)$\;}
\Modified{$T_i \cup \{ t_{ir}(1) \} \cup Y(t_{ir}(-g_{ic}g_{rc}^{-1}) w,t_{ir}(-g_{ic}g_{rc}^{-1}) u_1,u_2)$\;}
\BlankLine
\For{$\ell=0,\ldots,f-1$}{
compute $t_{i(i-1)}(\omega^\ell) = vt_{(i-1)(i-2)}(\omega^\ell)v^{-1}$\;
}
compute $t_{i(i-1)}(-g_{ic}g_{rc}^{-1})$\;
compute $t_{ir}(-g_{ic}g_{rc}^{-1}) = [t_{i(i-1)}(-g_{ic}g_{rc}^{-1}),t_{(i-1)r}(1)]$\;
compute $t_{ir}(1) = [t_{i(i-1)}(1),t_{(i-1)r}(1)]$\;
$w := t_{ir}(-g_{ic}g_{rc}^{-1}) w$\; 
$u_1 := t_{ir}(-g_{ic}g_{rc}^{-1}) u_1$\;
\end{algorithm}

At this point in each pass of the main `for' loop of Algorithm~\ref{alg:TaylorFull}, we call the subroutine \textsc{LeftUpdate}[$i$] for $i=r+2,\ldots,d$, unless $r \geq d-1$, in which case the current column will have already been cleared. 
The role of this subroutine is to effect the elementary row operations necessary to clear the rest of the current column (one entry of $g$ is made zero with each call to the subroutine). 
The cost of an MSLP for Algorithm~\ref{alg:ShiftTransvections} is as follows.

\begin{La} \label{la3}
Let $\lambda=2\log_2(q)+3f+10$ and $b=f+18$. 
There exists a $b$-MSLP, $S$, of length at most $\lambda$ such that if $S$ is evaluated with memory containing the input of Algorithm~$\ref{alg:ShiftTransvections}$ then $S$ returns memory containing the output of Algorithm~$\ref{alg:ShiftTransvections}$.
\end{La}

\begin{Prf}
The `for' loop computes $T_i$ in $2f$ instructions, and does not increase the input memory requirement of $f+1+|Y|=f+14$ slots because $T_i$ can overwrite $T_{i-1}$. 
With $-g_{ic}g_{rc}^{-1}$ assumed given as a polynomial in $\omega$, the transvection $t_{i(i-1)}(-g_{ic}g_{rc}^{-1})$ is constructed from $T_i$ in at most $2\log_2(q)+f$ matrix operations, with an additional memory requirement of three slots plus the one slot required to store $t_{i(i-1)}(-g_{ic}g_{rc}^{-1})$ for the next step. 
So, by this point, the subroutine has needed at most
\begin{equation} \label{mem4}
(f+14) + 3 + 1 = f+18
\end{equation} 
memory slots at any one time.
The transvection $t_{ir}(-g_{ic}g_{rc}^{-1})$ is formed by computing the commutator of $t_{i(i-1)}(-g_{ic}g_{rc}^{-1})$ and $t_{(i-1)r}(1)$. According to Section~\ref{ssec:ex}(i), this takes four instructions, but does not increase the memory requirement~\eqref{mem4} because one of the slots used to compute $t_{i(i-1)}(-g_{ic}g_{rc}^{-1})$ can now be overwritten while computing the commutator, which only requires one slot (plus the two inputs). 
The commutator $t_{ir}(1) = [t_{(i-1)r}(1),t_{i(i-1)}(1)]$ is then computed and replaces $t_{(i-1)r}(1)$ in memory, taking another four instructions but again not increasing the memory requirement~\eqref{mem4}.
Finally, the matrices $w$ and $u_1$ are multiplied by $t_{ir}(-g_{ic}g_{rc}^{-1})$ in two instructions without adding to~\eqref{mem4}.
So our claimed value for $b$ is given by~\eqref{mem4}, and our $b$-MSLP has length at most
$
2f + (2\log_2(q)+f) + 4 + 4 + 2 = 2\log_2(q) + 3f + 10.
$
\end{Prf}

Note that on the first pass of \textsc{LeftUpdate}[$i$], namely when $i=r+2$, the element $t_{(i-1)r}(1) = t_{(r+1)r}(1)$ is already contained in $T_{i-1}=T_{r+1}$. 
A copy of this element would be made at this point to form the required input.

\subsubsection{Calls to Algorithms~\ref{alg:LastTransvections} and~\ref{alg:BackshiftTransvections}}

\begin{algorithm}[!t] \label{alg:LastTransvections}
\caption{\textsc{LastTransvections}[$c$]}
\Input{$c \in \mathbb{N}$\;}
\Uses{$T_d \cup Y(w,u_1,u_2)$\;}
\Modified{$T_c \cup Y(w t_{c(c-1)}(-g_{r(c-1)}g_{rc}^{-1}),u_1,u_2 t_{c(c-1)}(-g_{r(c-1)}g_{rc}^{-1}))$\;}
\BlankLine
\If{$c\leq d-1$}{
\For{\textnormal{$\ell=0,\ldots,f-1$}}{
\For{\textnormal{$i=d-1,\ldots,c$}}{
compute $t_{i(i-1)}(\omega^\ell) = v^{-1}t_{(i+1)i}(\omega^\ell)v$\;
}
}
}
compute $t_{c(c-1)}(-g_{r(c-1)}g_{rc}^{-1})$\;
$w := w t_{c(c-1)}(-g_{r(c-1)}g_{rc}^{-1})$\;
$u_2 := u_2 t_{c(c-1)}(-g_{r(c-1)}g_{rc}^{-1})$\;
\end{algorithm}

\begin{algorithm}[!t] \label{alg:BackshiftTransvections}
\caption{\textsc{RightUpdate}[$j$]}
\Input{$j \in \mathbb{N}$\;}
\Uses{$T_{j+2} \cup \{ t_{c(j+1)}(1) \} \cup Y(w,u_1,u_2)$\;}
\Modified{$T_{j+1} \cup \{ t_{cj}(1) \} \cup Y(w t_{cj}(-g_{rj}g_{rc}^{-1}),u_1,u_2 t_{cj}(-g_{rj}g_{rc}^{-1}))$\;}
\BlankLine
\For{\textnormal{$\ell=0,\ldots,f-1$}}{
$t_{(j+1)j}(\omega^\ell) := v^{-1}t_{(j+2)(j+1)}(\omega^\ell)v$ for $\ell=0,\ldots,f-1$\;
}
compute $t_{(j+1)j}(-g_{rj}g_{rc}^{-1})$\;
$t_{cj}(-g_{rj}g_{rc}^{-1}) := [t_{c(j+1)}(1),t_{(j+1)j}(-g_{rj}g_{rc}^{-1})]$\;
$t_{cj}(1) := [t_{c(j+1)}(1),t_{(j+1)j}(1)]$\;
$w := w t_{cj}(-g_{rj}g_{rc}^{-1})$\; 
$u_2 := u_2 t_{cj}(-g_{rj}g_{rc}^{-1})$\;
\end{algorithm}

Once Algorithm~\ref{alg:ShiftTransvections} has been called the required number of times, the $c$th column of $g$ is clear and the main `for' loop of Algorithm~\ref{alg:TaylorFull} moves on to the row clearing stage for the $r(c)$th row. 
This is accomplished by Algorithms~\ref{alg:LastTransvections} and~\ref{alg:BackshiftTransvections}. Setting $r = r(c)$, the former makes the $(c-1,r)$-entry of $g$ zero by multiplying $g$ on the right by $t_{c(c-1)}(-g_{r(c-1)}g_{rc}^{-1})$, after first computing the transvections comprising $T_c$ (if $c\neq d$). 
The latter clears the rest of the $r$th row by multiplying $g$ by the appropriate transvections. 
The costs of MSLPs for these subroutines are evidently the same as for Algorithms~\ref{alg:FirstTransvections} and~\ref{alg:ShiftTransvections}, respectively. 
We summarise:

\begin{La} \label{la4}
Let $\lambda=f(2r-1) + 2\log_2(q) + 2$ and $b=f+17$. 
There exists a $b$-MSLP, $S$, of length at most $\lambda$ such that if $S$ is evaluated with memory containing the input of Algorithm~$\ref{alg:LastTransvections}$ then $S$ returns memory containing the output of Algorithm~$\ref{alg:LastTransvections}$.
\end{La}

\begin{La} \label{la5}
Let $\lambda=2\log_2(q)+3f+10$ and $b=f+18$. 
There exists a $b$-MSLP, $S$, of length at most $\lambda$ such that if $S$ is evaluated with memory containing the input of Algorithm~$\ref{alg:BackshiftTransvections}$ then $S$ returns memory containing the output of Algorithm~$\ref{alg:BackshiftTransvections}$.
\end{La}

\subsubsection{Total length and memory quota for Algorithm~\ref{alg:TaylorFull}}

The subroutine \textsc{FirstTransvections[$r$]} (Algorithm~\ref{alg:FirstTransvections}) is run whenever $r \neq 1$, namely $d-1$ times, with each value $r \in \{2,\ldots,d\}$ occuring exactly once (as explained earlier). 
According to Lemma~\ref{la2}, each call to \textsc{FirstTransvections[$r$]} contributes
\[
f(2r-1) + 2\log_2(q) + 2
\]
instructions to our MSLP for Algorithm~\ref{alg:TaylorFull}. 
Hence, in total, calls to this subroutine contribute
\begin{equation} \label{c1}
\sum_{r=2}^d (f(2r-1) + 2\log_2(q) + 2) = (d^2-1)f + (d-1)(2\log_2(q) + 2)
\end{equation}
instructions. 

For each column $c$ with $r = r(c) \leq d-2$, 
the subroutine \textsc{LeftUpdate[$i$]} is called for each $i$ such that $r+2 \leq i \leq d$. 
Each call to this subroutine requires $2\log_2(q) + 3f + 10$ MSLP instructions, according to Lemma~\ref{la3}. 
So 
\[
(d-r-1)(2\log_2(q) + 3f + 10)
\]
instructions are needed per column, yielding a total cost of
\begin{align} 
\nonumber
&\sum_{r=1}^{d-2} (d-r-1)(2\log_2(q) + 3f + 10) = \\
\label{c2}
& \qquad \qquad \frac{1}{2}(d-1)(d-2)(2\log_2(q) + 3f + 10)
\end{align}
instructions.

The total cost of all column clearing stages of Algorithm~\ref{alg:TaylorFull} is the sum of the expressions \eqref{c1} and \eqref{c2}. 
For simplicity, let us note that this sum is less than
\begin{equation} \label{c3}
d^2\left(\log_2(q) + \frac{5f}{2} + 5\right) + 2d(\log_2(q)+1).
\end{equation}
As noted above, the row clearing stages, that is to say calls to Algorithms~\ref{alg:LastTransvections} and~\ref{alg:BackshiftTransvections}, contribute (at most) as much as the column clearing stages. 
So {\em twice} the quantity \eqref{c3} is contributed to the maximum length of an MSLP for Algorithm~\ref{alg:TaylorFull}. 
In addition, we must also include the cost of the initial computation of $T_1$ given by Lemma~\ref{la1}, namely $5f-1$ instructions, and then two additional instructions in order to invert $u_1$ and $u_2$ before returning them (second- and third-last lines of Algorithm~\ref{alg:TaylorFull}). 
So we see that we can construct an MSLP for Algorithm~\ref{alg:TaylorFull} with length at most
\[
d^2(2\log_2(q) + 5f + 10) + 4d(\log_2(q) + 1) + (5f-1) + 2.
\]
The maximum memory quota $b$ for such an MSLP is just the maximum of all the $b$ values in Lemmas~\ref{la1}--\ref{la5}. 
This maximum is $b=f+18$, from both Lemma~\ref{la3} and Lemma~\ref{la5}.
This completes the proof of Proposition~\ref{dOddProp}.

\subsection{Modifications for $d$ even}

Let us now explain the changes required when $d$ is even. 
The main issue is that the formula~\eqref{eq:transRec} used to compute the sets of transvections $T_i$ recursively throughout our implementation of the algorithm described by Taylor looks {\em two} steps back instead of one when $d$ is even. 
That is, each $T_i$ is computed from either $T_{i-2}$ (while clearing a column) or $T_{i+2}$ (while clearing a row), not from $T_{i\pm 1}$ as was the case for $d$ odd. 
Therefore, an MSLP for the $d$ even case needs to hold {\em two} of the sets $T_i$ in memory at any given time, because a $T_i$ cannot be overwritten until until it has been used to compute $T_{i\pm 2}$. 
This adds (at most) $f$ memory slots to the maximum memory quota $b=f+18$ of Proposition~\ref{dOddProp}.

The other main change is that initially, namely for the first column clearing stage, the set of transvections $T_3$ must be computed from $T_2$ via the formula~\eqref{eq:t32}, whereas in the $d$ odd case it is computed via~\eqref{eq:transRec}. 
This adds only one extra MSLP instruction, in order to form and store the element $xv^{-1}$ needed in the conjugate on the right-hand side of~\eqref{eq:t32} (this element can later be overwritten and so does not add to the overall maximum memory quota; recall also that $x$ is no longer the identity when $d$ is odd). 
Observe that the formula~\eqref{eq:t21} differs from the $d$ odd case only in the sense that $v$ is replaced by $x^{-1}$, and hence the initial computation of $T_2$ requires the same number of instructions and memory slots as before. 
The sets $T_2$ and $T_3$ are computed as described above in preparation for the first column clearing stage, but are subsequently computed via the recursion~\eqref{eq:transRec} (with increased memory quota relative to the $d$ odd case, as already explained).

Although the described modifications are not complicated in and of themselves, they would introduce noticeable complications into our pseudocode and hence we have chosen to separate the $d$ even case for the sake of clearer exposition, opting to simply point out and explain the changes instead of writing them out in detail. 
Since the $d$ even case is slightly more costly than the $d$ odd case in terms of both the length (one extra instruction) and required memory ($f$ extra slots) for our MSLP, we conclude by extending Proposition~\ref{dOddProp} thusly:

\begin{Pro} \label{dBothProp}
Let $q=p^f$ with $p$ prime and $f \geq 1$, let $d$ be an integer with $d \geq 2$, and set $b=2f+18$ and
\[
\lambda=d^2(2\log_2(q) + 5f + 10) + 4d(\log_2(q) + 1) + (5f+2).
\]
There exists a $b$-MSLP, $S$, of length at most $\lambda$ such that if $S$ is evaluated with memory containing the list $Y(g,1,1)$ in~\eqref{eq:gens} then $S$ outputs memory containing $Y(w,u_1,u_2)$ with $g=u_1wu_2$ the Bruhat decomposition of~$g$.
\end{Pro}

\noindent {\em Proof of Theorem~$\ref{introThm}$.} Combing Proposition~\ref{dBothProp} with Propositions~\ref{Prop1} and~\ref{Prop2} yields the theorem: the maximum memory quota at any point is the $2f+18$ memory slots needed in Proposition~\ref{dBothProp}, while the lengths of the MSLPs in each case are $O(d^2\log(q))$. \hfill $\Box$

\section{Complexity of algorithm}  \label{sec6}
In addition to the length and number of memory slots of an MSLP, the complexity of computing the MSLP plays an important role in practice.   

\begin{Pro} \label{ComplexityOfAlgorithmMain}
Let $G = \SL(d,q)$ (in its natural representation) and $g \in G$. There exists an algorithm for computing an MSLP for writing $g$ in the LGO standard generators that has complexity $\mathcal{O}(d^3)$ measured in finite field operations of $\F_q$.
\end{Pro}

\begin{Prf}
The algorithm presented in this paper computes an MSLP as required. It is divided into three parts where an MSLP for the complexity of each part is independent of the complexity of the others. Computing an MSLP for the Bruhat decomposition corresponds to performing a special kind of Gaussian algorithm and has complexity $\mathcal{O}(d^3)$. Computing an MSLP for a permutation of $\operatorname{Sym}(d)$ requires $\mathcal{O}(d)$ operations. The complexity of computing an MSLP for a diagonal matrix is also $\mathcal{O}(d)$. Overall, this yields the claimed complexity of $\mathcal{O}(d^3 + d + d) = \mathcal{O}(d^3)$.
\end{Prf}

\begin{Rem}
\textnormal{
Note that this complexity for rewriting elements of $\SL$ in its natural representation is better than the complexity of previously known algorithms, e.g.\ Elliot Costi's algorithm \cite{Costi} which has complexity $\mathcal{O}(d^3 \log q)$ field operations.
}
\end{Rem}

\section{Implementation} \label{sec5}

Our algorithm has been implemented in the computer algebra system {\sf GAP}~\cite{GAP} and is available in the {\sf GAP} package {\sf BruhatDecomposition}~\cite{BruhatDecomposition}. We have tested our implementation on matrices of various sizes over finite fields.

For example, computing the Bruhat decomposition of a random matrix in $\GL(250,2)$ resulted in an SLP of length $353\;969$. During the evaluation, our MSLP required $32$ memory slots and it was easily possible to evaluate this MSLP on the standard generators of $\GL(250, 2)$. However, it was not possible to evaluate this SLP directly in {\sf GAP} by storing $353\;969$ matrices in $\GL(250, 2)$, as this required too much memory. (We remark that the number of memory slots used, 32, is slightly higher than the maximum of $2f+18=22$ slots asserted in Theorem~\ref{introThm}, because we stored a few extra group elements for convenience.)

We note that after applying the function \textsc{SlotUsagePattern}, the resulting SLP only required $12$ memory slots and could be evaluated in the same time as our MSLP. This is due to the fact that \textsc{SlotUsagePattern} was handed a well-designed SLP. When faced with an SLP not designed to be memory efficient, one might not expect such drastic improvements.

\section*{Acknowledgements}

We thank Max Neunh\"offer for his assistance and advice on the SLP functionality in {\sf GAP}. We thank Bill Unger for discussions and his advice on Magma implementations.

The first and last authors acknowledge funding by
 the SFB-TRR 195 ‘Symbolic Tools in Mathematics and their Application’ of the German Research Foundation (DFG) [Program ID 286237555].
The first and third authors acknowledge funding by the
Australian Research Council DP190100450.

\end{document}